\documentclass{elsarticle}
\usepackage{amssymb}
\usepackage{amsmath}
\usepackage{graphicx}
\usepackage{color}
\usepackage[usenames,dvipsnames]{xcolor}
\usepackage{multicol}
\usepackage{enumerate}
\usepackage{soul}

\usepackage[titletoc]{appendix}

\usepackage{pst-plot}
\usepackage{tikz}
\usetikzlibrary{arrows}
\usetikzlibrary{er,positioning}

\DeclareGraphicsExtensions{.pdf,.png,.jpg}

\font\manual=manfnt

\let\emptyset\varnothing

\newcommand{\eps}{\varepsilon}

\def\reals{\mathbb{R}}
\def\ireals{\mathbb{IR}}
\def\irepreals{\mathcal{X}}
\def\nat{\mathbb{N}}
\def\rat{\mathbb{Q}}
\def\integer{\mathbb{Z}}

\newcommand{\diam}{{\rm{diam}}\,}

\newcommand{\dom}[1]{{\rm dom}\,#1}

\newcommand{\inte}{{\rm Int}\,}

\def\strongcomp{$s$-IAC}
\def\weakcomp{$w$-IAC}

\def\qed{{\hfill{\vrule height5pt width3pt depth0pt}\medskip}}

\renewcommand{\phi}{\varphi}

\newtheorem{ex}{Example}
\newtheorem{theorem}{Theorem}
\newtheorem{remark}{Remark}

\newtheorem{definition}{Definition}
\newcommand{\proof}{{\bf Proof }}
\newtheorem{lemma}{Lemma}

\begin{document}

\begin{frontmatter}
\title{Real-number Computability from the Perspective of Computer Assisted Proofs  in Analysis}

\author{Ma{\l}gorzata Moczurad}
\ead{malgorzata.moczurad@ii.uj.edu.pl}

\author{Piotr Zgliczy{\'n}ski\corref{cor1}}\ead{piotr.zgliczynski@ii.uj.edu.pl}

\cortext[cor1]{Principial corresponding author}

\address{Jagiellonian University,
Faculty of Mathematics and Computer Science,\\
{\L}ojasiewicza 6, 30--348  Krak\'ow, Poland}

\begin{abstract}
Inspired by computer assisted proofs in analysis, we present an interval  approach to real-number computations.
\end{abstract}

\end{frontmatter}

\tableofcontents

\section{Introduction and motivation}\label{sec:intro}

There are various approaches to  the definition of computability and complexity of problems in continuous mathematics, e.g.\ numerically solving algebraic equations, ODEs or PDEs.  However  the standard numerical computations are not rigorous, as a finite computer is unable to represent the continuum. Depending on how they idealize  the non-rigorous computations performed on the existing computers,
various authors  (see for example~\cite{Aberth, KiK91, BCSS, Wei00, Bra05}) propose different approaches to the machine model used and to the notion of computability of objects.

We are not fully satisfied with these attempts, since we have specific requirements. Our point of departure are the existing computer assisted proofs  in analysis (we will use the acronym CAPA), like Lanford's proof  of the Feigenbaum conjectures for the period doubling cascade~\cite{La} or other
CAPAs in dynamics (see for example \cite{NeuRa93,NeuRaSch94,Tu99,Tu02,AK-10,MiMro-1995,MiMro-1998,BerzMakinoHofkens,ZglKsPer-2004,ZglRigAlg-2010}) and our personal experience with the CAPD library \cite{CAPD} for rigorous integration of ODEs (see also \cite{Ariadne,VNode,NedialkovJackson1998,MakinoBerz2006}).
Our goal is to obtain optimal algorithms for use in CAPAs. We would like to obtain the results in the spirit of IBC or BSS models, but without idealization of the computer.

In his book~\cite{Wei00} Weihrauch says:
\textit{Let us consider computable analysis as a mathematical theory of those real functions (...) which can be computed by physical machines like digital computers. Since we do not know the precise meaning of `computable by physical machines', every mathematical investigation in computable analysis must be based on a model of computation. Such a model of computation is not `true' or `false' but can merely be more or less realistic, powerful, expressive, illucidating or useful in practice according to the specific situation.}

This is also the case with this paper:
the model of computation presented
here is neither better nor worse than the other ones. It is a tool suitable
for certain applications but unsuited for others. Thus our model is
precisely {\em more [...] useful in practice according to the specific situation}: it is tailored for CAPA carried out in an interval
manner. The predominant computation in CAPAs is finding $f(K)$ with a given accuracy, where $f\colon \mathbb{R}^n \multimap\!\to \mathbb{R}^m$ is a partial function and $K$ is usually a compact set (although it might be a point).

\subsection{Intervals as a natural phenomenon} \label{subsec:nat-ph}
The majority of problems in analysis cannot be solved analytically, hence  approximate  numerical methods are developed. These methods usually give flawed solutions and sometimes give rise to misleading mathematical conclusions (see for example \cite{Hum1993}).

\begin{ex}\rm
\label{ex:ode-gen}
Consider an ODE  and the problem of  finding $x(t,x_0)$, a trajectory  starting from the given point
 $x_0\in \reals$ at time $t_0=0$.
Apart from very special cases this problem cannot be solved analytically and we are compelled to use numerical methods.
Even if we were able to represent $x_0$ and $t$ with an infinite precision as the actual real numbers, an inaccuracy would appear as a consequence of a method. The result we obtain is a set, for example, a product of intervals, which contains the error bound. This set becomes  the initial condition for the next time step. Therefore we need an arithmetic on sets (or intervals)   in real number computations with rigorous control of errors.
\mbox{}\hfill\manual\char'170
\end{ex}

Very often an input data $\tilde{x}$ comes from experiments thus they are not precise and we know only an upper bound for an error, i.e.\ an interval $[\tilde{x}-\eps, \tilde{x}+\eps]$. For many practical problems approximated results are insufficient and it is important to have  guaranteed estimates.  An example of interval computations of this type  can be found in \cite{BerzMakinoHofkens} , where a validated trajectory of a comet on the possible collision trajectory with the Earth and the Moon was computed and it was established that it would pass the Earth at a save distance. There  the uncertain positions of planets and major asteroids were given as interval valued parameters to a system of ODEs describing the comet motion. 
Observe also that in the present day computers with hardware built-in real arithmetic and efficient libraries for multiple precision arithmetic (see for example~\cite{GMP})
one has to deal with truncation errors, which immediately leads to interval arithmetic.

Using intervals to describe real numbers is essential for obtaining rigorous computations, where we can be sure that actual results belong to the computed intervals. To achieve this the following  strategy is used: the computed right-end of an interval should be a representable number not less than the actual right end, and analogously, the computed left end should be a representable number not greater than the actual left end. This is automatically realized by the interval arithmetic \cite{MKC09,Neumeier90}, see also Section~\ref{sec:intervals}.

\subsection{Notations}\label{subsec:notations}
For the ease of reference, we put major notations here; some of them may only become clear when they appear in their proper context in further sections.

Let $\Sigma$ be any finite and non-empty set. Define
\begin{eqnarray*}
\Sigma^0 & = & \{\eps\}\quad \mbox{(the language consisting only of the empty string)},\\
\Sigma^1 & = & \Sigma,\\
\Sigma^{n+1} & = & \{ wv : w \in \Sigma^i \mbox{ and } v \in \Sigma \}\quad \mbox{for each } i>0.
\end{eqnarray*}
We use $\Sigma^*$ to denote the usual Kleene closure, i.e.
$$\Sigma^{*}=\bigcup _{i\in \nat}\Sigma^i= \{\eps \}\cup \Sigma\cup \Sigma^2\cup \Sigma^3\cup \ldots .$$

We denote by $\integer$ the set of integer numbers, by $\rat$ the set of rational numbers and by $\mathbb{D}$ the set of {\em dyadic numbers}, i.e.\ numbers in the form $p/2^q$ where $p, q \in \integer$. Note that dyadic numbers correspond to floating point numbers: a floating point notation $m\cdot 2^{e}$ is equivalent to a dyadic number $m/2^{-e}$.

By $R$ we denote the set of representable numbers and by $\widehat{R}$ the set of all representations. We do not specify any particular set; we just assume some properties as pointed out in Section~\ref{sec:world}.
By ${\irepreals}$ we denote the set of all intervals with endpoints belonging to $\widehat{R}$ and we call them {\em representable intervals}.

 By $s_c$ we denote a step function with a discontinuity in point $c$, i.e. any function of the form
$$
s_c(x) = \left\{
\begin{array}{ll}
a_1, & x > c\\
a_2, & x \leqslant c,
\end{array}
\right.
$$
where $a_1, a_2\in \reals$. Note that it is not the values $a_1, a_2$ but just the position of discontinuity that is important in our
considerations.

\section{CAPAs in practice}\label{sec:capa}
From the point of view of CAPAs, it does not matter whether a real number is computable or not.
Each CAPA operates on representable numbers (usually some subset of rationals)  and on intervals with representable ends, obtaining properties valid for all real numbers (from some range of interest).

There is a lot of mathematically interesting questions, that can be reduced to a finite number of strict  inequalities between computable functions on some simple compacts in
$\mathbb{R}^n$.
Obviously in each CAPA   there is a substantial theoretical (i.e. non-algorithmic) work required to reduce it to the inequalities  to be checked rigorously.
To check a finite number of strict inequalities, finite precision computations are often sufficient, hence one can obtain true mathematical statements even for objects that might not be computable.

\begin{ex}\label{ex:brouwer}\rm
Given a computable continuous function $f:\reals^n\to \reals^n$ and a convex set $I$, we are interested whether there is a fixed-point of $f$ in $I$, i.e.\ $\exists x_0\in I:\ f(x_0) = x_0$.

By the Brouwer fixed-point theorem, if
$f(I) \subset I$, then there exists $x_0\in I$.
However we need stable (i.e. resistant to small perturbations) assumptions thus we require that $I$ is a ``simple'' set (e.g.\ a ball or a cube) and
$f(I)\subset \mbox{int}\, I$. It is obvious that the last condition can be formulated in terms of strong inequalities.

Notice that the computations are carried out on representable numbers, whilst the fixed-point need not be computable~\cite{Bai85}.
\hfill\manual\char'170
\end{ex}

\begin{ex}[see \cite{ZglLohner-2002,Lohner1,Lohner2}]\rm This is a continuation of Example~\ref{ex:ode-gen}, 
the rigorous integration of autonomous ODEs.
Assume we have  the  following initial value problem in $\reals^n$:
$$
\left\{
\begin{array}{lcl}
\stackrel{.}{x} & = & f(x)\\
x(0) & = & x_0\in \reals^n,
\end{array}
\right.
$$
where $f\in C^{\infty}$.
We start with $Z = [x_0, x_0]$.
Each time step of the rigorous  Lohner
integration algorithm has two stages. In the first stage we choose a set $W$ (using a heuristic procedure), such that:
\begin{itemize}
\item $Z\subset W$
\item a solution starting from a point in $Z$ exists in $W$ for $t\in [0,h]$, i.e.\ we check the condition $\varphi([0,h],Z)\subset W$, which can be reduced to a finite number of inequalities.
\end{itemize}
 In the second stage the solution after the time $h$ is obtained by evaluation in interval arithmetic of the Taylor formula on Z;  remainder term needs to be evaluated on $W$. The sum of these two intervals is the new value of $Z$.
\hfill\manual\char'170
\end{ex}

The use of interval arithmetic (see Section~\ref{sec:intervals}) is crucial to CAPAs which are not based on symbolic computations as it takes care of the round-off errors and allows evaluations of functions on set arguments.
This also provides a different kind of information in the analysis of the complexity of problems; in \cite{KiK91} or \cite{IBC1988, TW92} only values of a function (or its derivatives) at a point are allowed.\footnote{It is also possible to have a CAPA without using the interval arithmetic, i.e. performing all computations
in a finite precision arithmetic and then running a separate program (or doing pen and paper calculations) to estimate the influence of the round-off errors. This is the case of the work \cite{MiMro-1995,MiMro-1998}.  This approach does not differ conceptually from the one based on the systematic use of the interval arithmetic.}

In each approach to computability and complexity of problems in analysis there is always  an attempt to define the notion of a computable mathematical object,
for example a number, function, set etc. These definitions are different in various approaches.
In the present work  we adopt the following principle:
\begin{center}
\begin{tabular}{p{14cm}}
\em
The  object is \emph{computable} in the CAPA world, if we can verify various mathematical statements
about it using the computer rigorously.
\end{tabular}
\end{center}
It often happens that   some properties of mathematical objects are amenable to a CAPA and others are not. Therefore
it makes sense to discuss the computability up to a certain degree and the difference between computable and non-computable is blurred.

\subsection{Interval extension of a function}
\begin{definition}\label{def:hull}
An {\em interval hull} of $S\subset \reals$ is the smallest closed interval containing $S$, i.e.
$$\mbox{\rm hull}(S) = [\inf_{x\in S} x, \sup_{x\in S} x].$$
\end{definition}

A function $F:\ireals  \supset \dom(F) \to \ireals$  is called {\em an interval function}, where $\ireals$ denotes the set of all intervals with real endpoints.

\begin{definition}\label{def:fun-ext}
Let $f:\reals\to \reals$.
Any interval function $F$ such that $f(I)\subset F(I)$, for any $I\in \ireals$ is called an {\em interval extension of $f$}.
\end{definition}

The interval arithmetic, by its design, automatically computes an interval extension  of rational functions. For elementary functions (e.g.\ $\exp$, $\sin$, $\cos$, $\log$) there are appropriate extensions in a typical library for interval arithmetic.  The simplest way to compute an interval extension of $f$ is to replace all real variables in $f$ by the interval ones and executing the interval arithmetic evaluation of the expression defining $f$. This straightforward approach suffers from a serious drawback, the so-called \emph{dependency problem}: it might significantly overestimate the result.
The example below illustrates this phenomenon.

\begin{ex}\rm
We want to estimate the range of $f(x) = e^{-x}$ for $I = [0,h]$, where $h >0$ is a representable number. We consider a series expansion for $f(x)$, i.e.
$$e^{-x} = 1 - x + \frac{x^2}{2!} - \frac{x^3}{3!} + \ldots$$
Let us denote by $\widehat{\exp}$  the interval arithmetic realization of a `naive' summation of the above series term by term.

It is easy to see that
\begin{eqnarray*}
\sup \widehat{\exp}(-[0,h]) & \geqslant & 1 + \frac{h^2}{2!} + \frac{h^4}{4!} + \ldots = \cosh h = \frac{e^h + e^{-h}}{2},\\
\inf \widehat{\exp}(-[0,h]) & \leqslant & 1 - h - \frac{h^3}{3!} - \ldots = 1 - \sinh h = 1 - \frac{e^h - e^{-h}}{2}.
\end{eqnarray*}
Observe that we have  inequalities because of possible truncation errors.

Therefore we obtain
\begin{equation}\label{eq:exp-eval-diam}
  \left[ 1 - \frac{e^h - e^{-h}}{2}, \frac{e^h + e^{-h}}{2} \right] \subset \widehat{\exp}(-[0,h]).
\end{equation}
We have for $h \to 0$
\begin{equation}\label{eq:exp-diam}
  \diam\left(\widehat{\exp}(-[0,h])\right) \geqslant e^h - 1=h + \frac{h^2}{2!} + O(h^3) ,\quad \diam\left(e^{-[0,h]}\right)=1-e^{-h}=h-\frac{h^2}{2!}+ O(h^3),
\end{equation}
 hence
 \begin{equation}\label{eq:exp-diff-eval}
   \diam\left(\widehat{\exp}(-[0,h])\right) - \diam\left(e^{-[0,h]}\right)\geqslant h^2 + O(h^3).
 \end{equation}
Thus to estimate effectively the range of $e^{-x}$ on $I=[0,h]$, with $h$ large, it is not enough to take   $\widehat{\mbox{exp}}(-I)$, since interval arithmetic gives us a highly overestimated result: $\diam(\widehat{\mbox{exp}}(-[0,h])) \geqslant e^h - 1$ and it tends to infinity with $h\to\infty$. Fortunately, the overestimation of the diameter of the result is $O(h^2)$ with $h\to 0$, if we disregard the truncation errors.  This observation is the basic idea behind the design of the binary subdivision algorithm in Section~\ref{sec:subdiv-eval}.
\hfill\manual\char'170
\end{ex}

\begin{definition}
An interval extension $F$ of $f$ such that
$$F(X) = \mbox{\rm hull}(\{y\ |\ \exists x\in X: f(x) = y\}),$$
where $X\subset \ireals$ is called a {\em minimal interval extension}.\footnote{Note that although for any function $f:\reals \to \reals$ there exists a  minimal interval extension function $F$, the converse is not true. For example there is no $f:\reals\to\reals$ such that the constant interval function $F(X) = [-1,1]$ is its minimal extension.}

\end{definition}
Finding the minimal interval extension is computationally difficult:
even in the class of multi-variate polynomials the general problem is NP-hard.  When we fix the number of variables in a polynomial, a polynomial-time algorithm is possible,
but it requires too much computation time to be practical.\footnote{Following~\cite{KLRK}:
By the $\eps$-approximate basic problem of interval computations, we mean the following problem:-
Given $n$ rational intervals $X_i$, a computable continuous function $f$ that transforms $n$ real numbers $x_1, \ldots, x_n$ into a real number $f(x_1, \ldots, x_n)$ and a rational $\eps >0$, compute rational numbers $\widetilde{\underline{Y}}$ and $\widetilde{\overline{Y}}$ that are $\eps$-close to  the range's end-points, i.e.\ for which
$|\widetilde{\underline{Y}} - \underline{Y}|\leqslant \eps$ and
$|\widetilde{\overline{Y}} - \overline{Y}|\leqslant \eps$, where
$Y = [\underline{Y}, \overline{Y}] = f(X_1, \ldots, X_n)$.

{\em Theorem 3.1}\; For every $\eps > 0$, the $\eps$-approximate
basic problem of interval computations is NP-hard even for polynomial functions with rational coefficients.

{\em Theorem 4.1}\; For every $n$, there exists a polynomial-time algorithm
that, for any $\eps > 0$, for any polynomial $f(x_1, \ldots , x_n )$, and for any $n$
intervals $[\underline{x_i}, \overline{x_i}]$, computes $\eps$-approximations to the endpoints of the range
$f([\underline{x_1}, \overline{x_1}], \ldots, [\underline{x_n}, \overline{x_n}])$.
}
Fortunately, in the vast majority of CAPA applications we do not need a minimal extension; it is sufficient to have a reasonable\footnote{We say a {\em reasonable interval extension} of a function meaning {\em an interval extension returned by some interval arithmetic library}. Since there exist different libraries, using different algorithms and optimizations, there is no formal definition of this term.} interval extension.

%---------------------------------------------------
\subsection{Interval algorithms}\label{sec:int-alg}
 In this section we compare interval approach to real-number computability with traditional, non-interval ones. We want to emphasize two issues important from the point of view of CAPA:
\begin{enumerate}
  \item Information about the value of the function in chosen points is not the only accessible information; we can ask for an estimate of $f(I)$, i.e. a set $J$, such that $f(I) \subset J$. For most of the applications the condition below is satisfactory:
\begin{equation}\label{eqn:ext-quality}
f(I) \subset J\subset B(f(I),\eta(I)),
\end{equation}
where $\eta(I)$ is small (in finite precision) or tends to 0 (in multiple precision arithmetic), if $\diam(I)$ tends to 0. We can think about $\eta$ as a function characterizing the quality of an algorithm for the computation of $f$.

  \item For many applications, there is no need for a priori global information, like the knowledge of the Lipschitz constant or the modulus of continuity. Of course, the lack of global information affects an error estimation and complexity of the algorithm, but very often in CAPA such an approach is good enough; see Section~\ref{sec:trap-int} for a simple example and \cite{BerzMakinoHofkens} for some practically relevant computation of the comet's trajectory.
\end{enumerate}

\subsubsection{Integration of continuous functions using the trapezoidal rule}\label{sec:trap-int}

Let us compare two algorithms: the algorithm for the trapezoidal rule (TRA) with the uniform grid and its interval version (ITRA). Notice that both have the same  structure, but they differ on the objects they operate. TRA manipulates functions and numbers, whilst ITRA interval extensions and intervals. TRA returns a number being an approximation of $\int_a^b f(x)dx$, whilst ITRA returns an interval $J_n$ such that
$\int_a^b f(x)dx\in J_n$, for any $n\in\nat$. If we have arithmetic with infinite precision, then $\lim_{n\to\infty} I_n = \int_a^b f(x) dx$ and $\lim_{n\to\infty}\diam(J_n) = 0$.

$
\begin{array}{p{7cm}p{7cm}}
\mbox{\bf TRA} & \mbox{\bf ITRA} \\
{\em Input:} & {\em Input:}\\
$f$ -- a function to be integrated, & $F$ -- an interval extensions of a function to be integrated,\\
$[a,b]$ -- an interval of integration, & $[a,b]$ -- an interval of integration,\\
$n$ -- the number of grid points, & $n$ -- the number of grid points\\
\\
{\em Algorithm:} & {\em Algorithm:}\\
State $h = (b-a)/n$. & State $h = (b-a)/n$.\\
Define the partition $x_0, x_1, \ldots, x_n$ of $[a,b]$, such that $x_i = a + i h$, for $i = 0, \ldots, n$. & Define the partition $x_0, x_1, \ldots, x_n$ of $[a,b]$, such that $x_i = a + i h$, for $i = 0, \ldots, n$.\\
Compute
$I_n = h\sum _{i=1}^{n}{\frac {f(x_{i-1})+f(x_{i})}{2}}.$ & Compute $J_n = h\sum_{i = 1}^n F([x_{i-1}, x_i])$.\\
\\
{\em Output:} $I_n$ & {\em Output:} $J_n$

\end{array}
$

If we do not know the Lipschitz constant, then in both cases the result of the algorithm can be far from the exact solution. But the fundamental difference between those approaches is that in the case of ITRA we know the error bounds, whilst in TRA we do not:
\begin{enumerate}
\item Observe that $e = \diam(J_n)/2$ is an error of ITRA, thou without additional knowledge of the function $f$, i.e.\ the quality of the interval extension $F$ we cannot predict the size of $e$. However,
let us stress again that very often for CAPA such an information is fully sufficient.

\item
If we know the Lipschitz constant $L$ of $f$, the error of TRA  is $\frac{(b-a)^2 L}{2n}.$
\end{enumerate}

\subsubsection{Newton method for finding zeros of a function}

\subsubsection*{Newton-Kantorovich method}

Let $f:\mathbb{R}^n \to \mathbb{R}^n$  be a $\mathcal{C}^2$ function. 
The task is to find $\bar{x}$, such that
\begin{equation}\label{eq:Nm-to-solve}
  f(\bar{x})=0.
\end{equation}
For this end one consider the Newton method defined by
\begin{equation}\label{eq:NewtonMethod}
x_{k+1}=N(x_k) = x_k - Df(x_k)^{-1}\cdot f(x_k), \quad k=0,1,2,\dots
\end{equation}

The criterion for convergence of the sequence defined by (\ref{eq:NewtonMethod}) is given by the following Kantorovich Theorem~\cite{Polyak2006}
(we write its finite dimensional version)
\begin{theorem}\label{thm:Kantorovich}
Assume that $f$ is defined and twice continuously differentiable on a closed ball $\overline{B}(x_0,r)$,
the linear operator $Df(x_0)$ is invertible, $\|Df(x_0)^{-1}f(x_0)\| \leqslant \eta$, $\|Df(x_0)^{-1}D^2f(x)\| \leqslant  K$, for all $x \in  \overline{B}(x_0,r)$, and
\begin{equation}
h = K\eta < \frac{1}{2}, \quad  r \geqslant\frac{1-\sqrt{1-2h}}{h} \eta.  \label{eq:Kantorovich-cond}
\end{equation}
Then Eq.~(\ref{eq:Nm-to-solve}) has a unique solution $\bar{x}  \in \overline{B}(x_0,r)$, the process (\ref{eq:NewtonMethod}) is well defined and converges to $\bar{x}$ with quadratic rate:
\begin{equation}
\|x_k - \bar{x}\| \leqslant \frac{\eta}{h 2^k} (2h)^{2k}.
\end{equation}
\end{theorem}

Using the above theorem we can  formulate the stopping criterion for iteration~(\ref{eq:NewtonMethod}). Let us fix $\eps>0$. We set $$K=\sup_{x \in \overline{B}(x_k,\eps)} \|Df(x_k)^{-1}D^2f(x)\|,\quad \eta=\|Df(x_k)^{-1}f(x_k)\|$$ and check whether the following conditions are satisfied (cf.~(\ref{eq:Kantorovich-cond}))
\begin{equation} \label{eq:N-stop}
 K\eta < \frac{1}{2}, \quad  r =\frac{1-\sqrt{1-2K \eta}}{K} \leqslant \eps.
\end{equation}
The test $r \leqslant \eps$ is needed to make sure that the constant $K$ has been computed over a set containing $\overline{B}(x_k,r)$ as required
in Theorem~\ref{thm:Kantorovich}. If (\ref{eq:N-stop}) holds, then we know that there is a unique solution of (\ref{eq:Nm-to-solve}) in
$\overline{B}\left(x_k, r\right)$. 
Somewhat relaxed version (non-rigorous one) of the stopping condition would be to set $K=\|Df(x_k)^{-1}D^2f(x_k)\|$. One would expect that if $\eps$ is small, then this choice of $K$ is close to the true value.

\subsubsection*{Interval Newton method}

The following theorem by Moore is the basis for the interval Newton method.
\begin{theorem}\cite{MKC09}
\label{thm:newton} Let $f:{\mathbb R}^n \to {\mathbb R}^n$ be a
$\mathcal{C}^1$ function. Let $X=\Pi_{i=1}^n[a_i,b_i]$, $a_i < b_i$. Assume
the interval enclosure of $Df(X)$, denoted here by $[Df(X)]$, is
invertible. Let $x_0 \in X$ and we define
\begin{equation}
  N(x_0,X) = - [Df(X)]^{-1}f(x_0) + x_0
\end{equation}
Then
\begin{itemize}
  \item if \ $x_1,x_2 \in X$ and $f(x_1)=f(x_2)$, then $x_1=x_2$;
  \item if \ $N(x_0,X) \subset X$, then $\exists !\ x^* \in X$ such that
   $f(x^*)=0$;
  \item if \ $x_1 \in X$ and $f(x_1)=0$, then $x_1 \in N(x_0,X)$;
  \item if \ $N(x_0,X) \cap X = \emptyset$, then $f(x) \neq 0$ for all $x \in X$.
\end{itemize}
\end{theorem}

In view of the above theorem one can devise the following iteration scheme
\begin{equation}
  X_{k+1}=X_{k} \cap N(x_k,X_k), \quad \mbox{$x_k$ is any point in $X_k$}.  \label{eq:NM-interval}
\end{equation}
where $X_0$ is a given interval set in which we look for solution of (\ref{eq:Nm-to-solve}) or we would like to rule out the existence of solution.

More precise formulation of iteration (\ref{eq:NM-interval}) is as follows.
\begin{enumerate}
\item Let $x_k$ be a mid point of $X_k$, we compute $Y=N(x_k,X_k)$. This may fail, if $[Df(X_k)]$ is not invertible. In this case we interrupt the iteration and return \textbf{failure}
\item If   $Y \subset X_k$, then in $X_k$ contains exactly one solution of (\ref{eq:Nm-to-solve}). We interrupt the iteration and return \textbf{solution-found} and $Y$ is the bound for the solution of (\ref{eq:Nm-to-solve}). In fact this also proves that there is a unique
    solution of (\ref{eq:Nm-to-solve}) in $X_0$.
\item If $Y \cap X_k =\emptyset$, then  we interrupt the iteration and return \textbf{there is no solution in $X_0$}
\item Otherwise we set $X_{k+1}=X_k \cap Y$ and go to the first step
\end{enumerate}
Observe that if in the second step we obtain \textbf{solution-found}, then we can continue with iteration (\ref{eq:NM-interval}) as long as we want to obtain tighter bounds for the unique solution of (\ref{eq:Nm-to-solve}) in $X_0$ (in a finite precision interval arithmetics $X_k$ will stabilize).

\subsubsection*{Comparison}
Let us present an ordinary (NM) and an interval (INM) versions of Newton algorithms. In both approaches an initial guess is needed which requires some preparation beforehand. Hence we give this value as parameter $x$ or $X$, respectively.

%--------------------------------------
$
\begin{array}{p{7cm}p{7cm}}
\mbox{\bf NM} & \mbox{\bf INM} \\
{\em Input:} & {\em Input:}\\
$f$ -- a function, & $F$ -- an interval extensions of a function,\\
$x$ -- an initial point, & $X$ -- an initial interval\\
$\eps$ -- desired precision &  \\
\\
{\em Algorithm:} & {\em Algorithm:}\\
{\bf do forever} \{ & {\bf do forever} \{\\
  & \mbox{}\hspace*{3mm} $\displaystyle{x = \frac{\mbox{\rm right}(X)+\mbox{\rm left}(X)}{2}}$;\\
\mbox{}\hspace*{3mm} {\bf if } $Df(x)$ is not invertible {\bf then}   & \mbox{}\hspace*{3mm} {\bf if } $[Df(X)]$ is not invertible  {\bf then} \\
\mbox{}\hspace*{6mm}  {\bf failure}; & \mbox{}\hspace*{6mm} {\bf failure};\\
\mbox{}\hspace*{3mm} $y = x - Df(x)^{-1}\cdot f(x)$; & \mbox{}\hspace*{3mm} $Y = x -[Df(X)]^{-1}f(x)$; \\
\mbox{}\hspace*{3mm} $K = \|Df(y)^{-1} D^2f(y) \|$; & \\
\mbox{}\hspace*{3mm} $\eta = \|Df(y)^{-1}f(y) \|$; & \\
\mbox{}\hspace*{3mm} {\bf if } $K\eta < \frac{1}{2}$ {\bf then} $\{$\\
\mbox{}\hspace*{6mm} $r = \frac{1 - \sqrt{1 - 2 K\eta}}{K}$;\\ 
\mbox{}\hspace*{6mm} {\bf if } $r\leqslant \eps$ {\bf then}
 {\bf solution-found} & \mbox{}\hspace*{3mm} {\bf if } $Y\subset X$ {\bf then } {\bf solution-found};\\
 \mbox{}\hspace*{3mm} \}\\
 & \mbox{}\hspace*{3mm} {\bf if } $Y\cap X = \emptyset$ {\bf then } {\bf no-solution};\\
\mbox{}\hspace*{3mm} x = y & \mbox{}\hspace*{3mm} $X = X\cap Y$\\
\} & \}\\
\\
{\em Output:} $y$ or {\bf failure} & {\em Output:} $Y$ or {\bf failure} or {\bf no-solution}
\end{array}
$

%--------------------------------------
\bigskip

Let us emphasize some important features of the above methods.
\begin{itemize}
\item In the interval Newton approach the initial set $X_0$  does not need to be  small, the only assumption is the invertibility of $[Df(X_0)]$.
\item When using the Newton iteration (\ref{eq:NewtonMethod}) the point $x_k$ wanders through the space until it eventually enters the region 
  where the assumption  of Theorem~\ref{thm:Kantorovich} is satisfied.
\item For the interval Newton method, at each step of the iteration we know that if there is a zero of $f$ in $X_0$, then it belongs to $X_k$. We need \emph{no global information} to obtain a rigorous result.

\item To obtain an error bound for the Newton iteration (\ref{eq:NewtonMethod}), we need to compute $K$ which involves the estimation of $D^2f(x)$ over a set. This can be done in two ways:
  \begin{itemize}
    \item {\bf using interval computations:}  compute $D^2 f (\overline{B}(x_k,\eps))$ `directly,' or
    \item {\bf using global information:} we need to know  a constant $M=\sup_{x \in U}\|D^2 f(x)\|$, where $U$ is some set of macroscopic size, containing the region of interest. Then $K=\|Df(x_k)^{-1}\| \cdot  M$. Observe that the stopping criterion should also include the test $\overline{B}(x_k,\eps) \subset U$.
  \end{itemize}
\end{itemize}

\subsubsection{Binary subdivision algorithm for the range evaluation}
\label{sec:subdiv-eval}

Here we describe a binary subdivision algorithm (BSA). As an example of use of BSA, we chose the problem of evaluation of the range of a function with desired accuracy.
However, BSA might be used in many applications, for example to compute $\int_I f(x)dx$, $\sup_{x\in I} f(x)$ or $\inf_{x\in I} f(x)$ with desired precision. Note that, in contrast to previous examples, there is no simple non-interval counterpart for that algorithm.

The algorithm runs for any function $f:\mathbb{R} \multimap\!\to \mathbb{R}$.
The input consists of:
\begin{itemize}
\item the interval function $F$\footnote{A function $f$ is given as a sub-procedure $F$ for an interval machine implementing the BSA.}, such that $f(J) \subset F(J)$, for any $J$;
\item $I$, the interval for which we estimate the range of $f$;
\item an accuracy bound $\eps > 0$, which for an interval machine has to be an interval, i.e. $E = [\eps, \eps]$.
\end{itemize}
$F$ may not be defined for some intervals, for example when dividing by an interval containing zero or when diameter of an interval exceeds the local convergence radius for an analytic function. However, we
assume that the Turing Machine computing $F(I)$ always stops and decides whether the computation is successful and $F(I)$ is defined.

%%%%%%%%%%%%%%%%%%%%%%%%%%%%%%%%%%%%%%%%%%%%%%%%%%%%%%%%%%

Here is the algorithm:
\begin{itemize}
\item[] {\em Input:} $F$, $I=[a, b]$, $E = [\eps,\eps]$

\begin{enumerate}
\item[1.] $S := \emptyset$ (an empty interval).

\item[2.] Define the initial partition of $[a,b]$ by setting $[a_0,a_1]=[a,b]$ and mark this interval as \textbf{bad}.

\item[3.]
Assume that $[a,b]$ has already been partitioned into $m$ intervals, i.e.,
$$[a,b]:= [a_0, a_1]\cup [a_1, a_2]\cup\ldots\cup [a_{m-1}, a_m]$$
and there exists $J\subset \{1, \ldots, m\}$ such that  $[a_{i-1},a_i]$ is \textbf{bad} for $i \in J$, while other intervals  are  \textbf{proper}.
Let $K=[a_{k-1},a_k]$ be  the longest \textbf{bad} interval.

If $K$ cannot be bisected (this may happen when $\diam(K)=0$ or the midpoint of $K$ is not a representable number), then stop and
return \textbf{failure}.

Otherwise:
\begin{enumerate}
\item bisect $K$ into $K_L$ and $K_R$,
\item remove $K$ from the partition and add both $K_L$ and $K_R$,
\item for $Z=K_L$ and $Z=K_R$ do the following:
\begin{itemize}
  \item compute $F(Z)$,
  \item if $F(Z)$ is defined and  $\diam(F(Z)) \leqslant \eps$, then $Z$ is \textbf{proper} and set $S = \mbox{hull}(S\cup F(Z))$, else $Z$ is \textbf{bad}.
\end{itemize}
\end{enumerate}

Repeat step 3 as long as the set $J$ of \textbf{bad} intervals is not empty.

\end{enumerate}

\item[] {\em Output:} $S$.
\end{itemize}

\begin{ex}\rm
Consider the function
$$
s_c(x) = \left\{
\begin{array}{ll}
a_1, & x > c\\
a_2, & x \leqslant c,
\end{array}
\right.
$$
where $c, a_1, a_2\in \reals$, $a_1 < a_2$. For any $a, b\in\reals$, $a<b$ and $\delta \geqslant 0$, let us denote
$$\mbox{rh}_\delta([a,b]) = [\underline{a}, \overline{b}],$$
such that $\underline{a}$, $\overline{b}$ are any but fixed representable numbers (see Section~\ref{sec:world}) such that $a - \underline{a} < \eps$ and $\overline{b} - b<\eps$.
For any $\delta \geqslant 0$ the function
\begin{eqnarray*}
F_\delta(I) & \equiv & [\mbox{if } (I\leqslant \mbox{rh}_\delta([c,c])) \mbox{ then } \mbox{rh}_\delta([a_2, a_2])\\ & &
\mbox{ else } \{
\mbox{if } (I > \mbox{rh}_\delta([c,c]))\ \mbox{then}\ \mbox{rh}_\delta([a_1,a_1])\\
  & & \quad\quad~
\mbox{ else }\ \mbox{rh}_\delta([a_1, a_2])\}].
\end{eqnarray*}
is the interval extension of $f$. Note that only when all numbers $c, a_1, a_2$ are representable then there exists computable minimal extension of $f$ and it is $F_0$.
If we have an infinite precision arithmetic, BSA never stops for $F_\delta$, when $\delta\neq 0$. On the contrary, if we have a finite precision arithmetic, the program always stops but returns failure when $\eps < \diam(\mbox{rh}_\delta([a_1, a_2]))$.
\hfill\manual\char'170
\end{ex}

\begin{remark}
If BSA stops and does not return failure then $f(I) \subset S \subset \overline{B}(f(I),\eps)$.
\end{remark}

Obviously, if we know more about a  function $f$ then the range can be computed faster using other approaches. For example for an  analytic function computable in polynomial time (with respect to $\log_2 (1/\eps)$) the maximum value on the compact interval can be computed
in polynomial time by searching for all roots of $f'(x)=0$ (see \cite[Sec. 6.2]{KiK91}).

\section{Real numbers and their representations}\label{sec:world}

Computers are discrete machines unable to represent the continuum of real numbers. Therefore they operate on a subset
of $\mathbb{R}$ called the set of representable numbers\footnote{It is a common approach to computability on real numbers to distinguish between computable and non-computable numbers. This very problem is not what we are interested in:
we do exact calculations on simple representations and infer properties of the entire sets.

Note that there is a significant difference in what we want to attain: whether we want to be able to compute (any digit of) a number, or to represent it (write down the entire number in some way). Thus  {\em computable numbers} are not the same as {\em representable numbers}.
However, there is no general definition of representable real numbers. The program (machine) computing a number can be  one of its possible representations.
A common representation is floating-point numbers with a finite mantissa of fixed length, which gives just a finite set of numbers (a subset of dyadic numbers).}, which  can be written as finite strings over a finite alphabet.  This assumption leads to a fundamental statement about a set of representable numbers.

\begin{remark}\label{rem-countable}
The set of representable numbers is  countable (finite or infinite).
\end{remark}

\begin{definition}
Let $\Sigma$ be a finite alphabet. We say that a function $r:\reals\to \mathcal{P}(\Sigma^*)$ is a {\em representation} if $$\forall x, y\in\reals: x\neq y\Longrightarrow r(x)\cap r(y) = \emptyset.$$
A {\em representation of a real number} $x$ with respect to $r$ is any element of the set $r(x)$. A number $x$ is {\em representable}  if $r(x)\neq\emptyset$.
\end{definition}

\begin{ex}\rm
In dyadic notation $1/2$ can have different representations as
$$\frac{1}{2},\quad \frac{2}{2^2},\quad \frac{4}{2^3},\quad \ldots$$
Thus the function $r$ is
$$r(x) =
\left\{
\begin{array}{ll}
\left\{\frac{p 2^k}{2^{n+k}}\ :\ k \in\nat \right\}, & \mbox{if } x = \frac{p}{2^q} \mbox{ with $p$ odd}\\
\\
\emptyset, & \mbox{otherwise}.
\end{array}
\right.
$$
\hfill\manual\char'170
\end{ex}
By $R_r$ we denote the set of representable numbers with respect to the representation $r$; then
$$\widehat{R}_r = \bigcup_{x\in\reals} r(x)$$
is a set of all representations with respect to $r$.

It does not make sense to consider the computability of $r$, but we still want to be able to decide whether a given string over $\Sigma$ represents a number or not.

We do not require that $R_r$ is closed under  arithmetic operations used since commonly used representations do not satisfy this requirement. For example, the dyadic number representation  is not closed under division, whilst floating point representation is closed neither under multiplication, nor division. Therefore we state a weaker condition.
\begin{definition}\label{def:faith}
Let $r$ be a representation.
Whenever we use an operation on representable numbers, say $f: R_r^{m}\to R_r$, it has to be {\em implemented faithfully}:
$$\forall \vec{x}\in R_r^{m}:\ r(f(\vec{x}))\neq\emptyset \Longrightarrow
\forall \vec{w}\in r(\vec{x}): \widehat{f}(\vec{w})\in r(f(\vec{x})),$$
where $\widehat{f}:\ \widehat{R}_r^m\to \widehat{R}_r$ is the implementation of $f$.
\end{definition}

\begin{remark}
From our point of view, $r$ is an {\em interesting representation} if it satisfies the following conditions:
\begin{enumerate}
\item $\forall w\in\Sigma^*$ it is decidable whether $w$ is a representation of some number $x$, i.e.\ if there exists $x$ such that $w\in r(x)$;

\item $\forall z\in\integer:\ r(z)\neq\emptyset$;
\item operations we want to use can be implemented faithfully (see Definition~\ref{def:faith}).
\end{enumerate}
\end{remark}

Since all results we present are independent of the specific set of representable numbers, further throughout the paper we do not specify a representation $r$ and  we just call the set of representable numbers $R$ and the set of representations $\widehat{R}$.
\begin{remark}\label{rem:conv}
Moreover, since in practice we usually identify (real) numbers with their representations, whenever  possible, we use this convention in the present paper: we say {\em representable real number} meaning a representation of that number.
\end{remark}
 To illustrate the ideas we often use the example of rational numbers or their finite subset. These two cases are important because the former is infinite and the latter corresponds to the floating point arithmetic realized in the present day computers.

\section{Computable functions }
\label{sec:comp-func-intv-machine}

The goal of this section is to discuss the notion of  a real function computable on the interval machine. The interval machine is a simple RAM-like model, equivalent to the Turing machine. It has one fundamental advantage though: it is equipped with an arithmetic which operates on intervals. This is absolutely crucial for CAPAs.  For the detailed definition of the interval machine see Appendix~\ref{def:machine}.

Let us recall that by $R$ we denote
the set of representable reals on which the interval arithmetic is built.
Denote by ${\mathcal{RI}}$ the {\em class of representable interval functions}, i.e.\ the class of functions computed by interval machines (see Definition~\ref{def:int-machine-comp}).
 For the following definitions we need the convention mentioned in Remark~\ref{rem:conv}: let $[\widehat{\alpha}, \widehat{\beta}]\in\irepreals$ and $f\colon \reals\to\reals$. By $f([\widehat{\alpha}, \widehat{\beta}])$ we mean the image of $f([\alpha, \beta])$, where $ [\alpha, \beta]\subset \reals$, $\widehat{\alpha}\in r(\alpha)$ and $\widehat{\beta}\in r(\beta)$.

We start with two definitions of a computable function. We distinguish between weak and  strong computability of functions; we abbreviate these terms as \weakcomp\ and \strongcomp, IAC standing for interval arithmetic computable.
Note that the weak computable functions are effectively (not necessarily polynomially) computable, i.e.\ there exist algorithms for interval machine  that compute them.
We also  distinguish between computability and $\eps$-computability. The former is stronger since we demand that some properties hold regardless of accuracy $\eps$; we denote it with an $\eps$ prefix: $\eps$\weakcomp\ and $\eps$\strongcomp.

\subsection{Strong interval arithmetic computable (\strongcomp) functions}
\label{subsec:strong-comp}

As it was mentioned in Section~\ref{sec:capa}, the crucial task in CAPA is computing the estimate of an image of a set with desired precision. In this section we formalize the class of functions for which it is possible to do this.
We define \strongcomp\ functions; this class  happens to be equivalent to the computable functions
in the sense of Definition~\ref{def:comp-func-ko} of Ko or Grzegorczyk recalled in Section~\ref{sec:othermodels}. Hence our approach does not affect the concept of commonly accepted definitions of computability in real numbers.

\begin{definition}\label{def:fun-eps-comp-new}
Let us fix $\eps>0$. We say that a function $f:\reals\supset \dom(f) \to\reals$ is {\em $\eps$\strongcomp} if there exists $ F \in \mathcal{RI}$ such that for any compact $K \in \irepreals$, $K \subset \dom(f)$ there exists an algorithm to establish a finite sequence $I_1, \ldots, I_n\in \irepreals$ such that
\begin{equation}
  K \subset \bigcup_{i=1}^{n} I_i\subset \dom(f) \quad\mbox{and}\quad \forall i\in\{1, \ldots, n\}:\ \left[ f(I_i)\subset F( I_i)\wedge
\diam(F( I_i)) \leqslant \eps\right].\label{eqn:e-comp-new}
\end{equation}

We say that $f$ is {\em \strongcomp} if there exists $ F \in \mathcal{RI}$ such that for any given $\eps>0$ and for any compact set  $K \in \irepreals$, $K \subset \dom(f)$ there exists an algorithm to establish  a finite sequence $I_1, \ldots, I_n\in\irepreals$ such that

\begin{equation}\label{eqn:comp-new}
K \subset \bigcup_{i=1}^{n} I_i\subset \dom(f) \quad\mbox{and}\quad \forall i\in\{1, \ldots, n\}:\ \left[ f(I_i)\subset F(\eps, I_i)\wedge
\diam(F(\eps, I_i)) \leqslant \eps\right].
\end{equation}
\end{definition}
\begin{remark}\label{rem:multi-dim-def1}
Definition~\ref{def:fun-eps-comp-new} carries over to multidimensional setting by demanding that $I_i$ become elementary multidimensional interval sets, for example products of representable intervals or some other simple sets.
\end{remark}

If $R=\mathbb{Q}$ (i.e. arithmetic with infinite precision), then it is easy to see that basic arithmetic operations ($+, \cdot$ and $1/x$), function $\sqrt{\mbox{\hspace*{1mm}}}$ and
many elementary analytic functions like $\exp$ or trigonometric functions are \strongcomp.
However in the case of $R$ being floating point numbers of finite size (this is the hardware standard in the present day computers), these operations are only $\eps$\strongcomp\ (on compact domains, with $\eps \to 0$ when the number of bits in the representation is increasing to infinity).

\begin{remark}\label{rem:BSA-inf}

BSA with infinite precision interval arithmetic  allows to compute any rational function on any compact
  interval with arbitrary precision.
  Therefore with such interval arithmetic these functions are \strongcomp\ in the sense of Definition~\ref{def:fun-eps-comp-new}.
  Analogous results can be easily proved for other elementary functions $\exp$, $\sin$, $\cos$ etc. and their compositions.

\end{remark}

The following lemma states that in conditions (\ref{eqn:e-comp-new}) and (\ref{eqn:comp-new}) we can assume all $I_i$'s to have positive
diameter, i.e. we can remove all degenerate intervals.
\begin{lemma}
\label{lem:remove-zero-len}
   Assume that $K\in \irepreals$ and $K \subset \bigcup_{i=1}^n I_i$, where $I_i \in \irepreals$. Then $K \subset \bigcup_{i \in Z} I_i$, where
    $i \in Z$ iff $\diam(I_i)>0$.
\end{lemma}
\proof
Let $K=[a,b]$. Assume that we have an interval $I_j$ with zero diameter, i.e. $I_j=[x,x]$. We would like to remove it from the covering of $K$.

We have two cases:
\begin{enumerate}
  \item There exists  $I_i$ such that $x \in \inte I_i$, then we can discard $I_j$ from the covering.
  \item For all intervals $I_i$, $x \notin \inte I_i$.

We argue that there exist  $c_1 <x < c_2$, such that the intervals $[c_1,x]$ and $[x,c_2]$ belong to the covering (if $x=a$ or $x=b$ then we just need one of these intervals).

Existence of the interval on the left: let us consider all intervals $I_i=[l_i,r_i]$ such that $l_i < x$. Obviously,
$r_i \leqslant x$, otherwise $x \in \inte I_i$. The largest of such $r_i$, say $r_{i_0}$, must be equal to $x$, otherwise the open interval
$(r_{i_0},x)$ will be not covered by $\bigcup I_i$.

The interval on the right is obtained analogously.
\qed
\end{enumerate}

\begin{theorem}
\label{thm:scmp-cont}
  Assume that $f:[a,b] \to \mathbb{R}$ is \strongcomp. Then $f$ is continuous.
\end{theorem}
\proof
Let us fix $\eps$ and let $[a,b]=\bigcup_{i=1}^n I_i$,  where $\{I_i\}$ are as in condition (\ref{eqn:comp-new}). For every $x \in [a,b]$ we set
 $$U_x=\bigcup \{I_i: i = 1, \ldots, n,\ x \in I_i\}.$$
 It is easy to see that $x \in \inte U_x$ (see proof of Lemma~\ref{lem:remove-zero-len}) and for all $y \in U_x$, $|f(y) - f(x)| \leqslant \eps$. Indeed if $y \in U_x$ then there is $I$, such that $y,x \in I$, hence
$$|f(x) - f(y)| \leqslant \diam(f(I)) \leqslant \diam(F(I)) \leqslant \eps.$$
\qed

 \begin{lemma}
\label{lem:cover}
   Assume that $K=[a,b]$ and $K \subset \bigcup_{i=1}^n I_i$, where $I_i=[l_i,r_i]$ with $l_i<r_i$. Let $\delta=\min_{i=1,\dots,n} \diam(I_i)$.

Assume $K=[a,b] \subset \bigcup_{i=1}^n I_i$, where $I_i=[l_i,r_i]$ with $l_i<r_i$. Let $\delta=\min_{i=1,\dots,n} \diam(I_i)$.
If $x,y \in K$ and $|x-y|\leqslant \delta$, then there exist $k,j$ such that $x \in I_j$, $y \in I_k$ and $I_j \cap I_k \neq \emptyset$.
\end{lemma}
\proof
If there exists $i$, such that $x,y \in I_i$, then take $j=k=i$.

Assume now that such $i$ does not exists.  Without loss of generality, assume that $x<y$.  Let $j$ be such that $r_j = \max\{r_i:\ i = 1, \ldots, n,\ x\in I_i\}.$
Obviously $r_j < y$, otherwise $x,y \in  I_j$.
Analogously, let $k$ be such that $l_k = \min\{l_i:\ i = 1, \ldots, n,\ y\in I_i\}.$
Obviously $l_k > x$, otherwise $x,y \in  I_k$.

We show that $r_j\geqslant l_k$ meaning that $I_j \cap I_k \neq \emptyset$. Assume the contrary. Let $z=(r_j + l_k)/2$. Consider $I_z=[l_z,r_z]$, such that $z \in I_z$. From the assumption, $x \notin I_z$ and $y \notin I_z$, hence $I_z \subset (x,y)$ and
$\diam(I_z) < \delta$, which contradicts the definition of $\delta$. Therefore $I_j \cap I_k \neq \emptyset$.
\qed

The next theorem shows that the information about the modulus of continuity can be obtained for  $\eps$\strongcomp\ functions.
\begin{theorem}
\label{thm:sccomp-unifrom-conf}
Let us fix any $\eps >0$. Assume that $f:[a,b] \to \mathbb{R}$ is  $\eps$\strongcomp,  i.e.\ there exists a finite covering of $[a,b]$ required by Definition~\ref{def:fun-eps-comp-new} of a nondegenerate interval (see Lemma~\ref{lem:remove-zero-len}) :
$[a,b]=\bigcup_{i=1}^n I_i$.
Let $\delta=\min_{i=1, \ldots, n} \diam(I_i)$.
Then for all $x,y \in [a,b]$ such that $|x-y|\leqslant \delta$, $|f(x) - f(y)|\leqslant 2 \eps$.
\end{theorem}

\proof
Let us take any $x,y\in [a,b]$, $|x-y|\leqslant \delta$. By Lemma~\ref{lem:cover}, there exist $I_i, I_j$ such that $x\in I_i$, $y\in I_j$ and $I_i\cap I_j\neq\emptyset$. Let us take any $z\in I_i\cap I_j$. Then
  $|f(x) - f(z)| \leqslant\eps$ and  $|f(y) - f(z)| \leqslant \eps$, therefore $|f(x)-f(y)| \leqslant 2 \eps$.
\qed

The above theorems are not surprising. In Ko's  approach  computable functions are continuous with computable modulus of continuity on compact domains.
 Below we show that  Definition~\ref{def:fun-eps-comp-new} is equivalent to the one given by Ko (see Section~\ref{sec:ko}), however, the complexity dependence remains an open problem.  

\begin{theorem}
\label{thm:sc-equiv-ko}
Assume we have an interval arithmetic with infinite precision. Function $f: K \to \mathbb{R}$, where $K \in \irepreals$, is \strongcomp\ iff it is computable in the sense of Ko, i.e. Definition~\ref{def:comp-func-ko}.
\end{theorem}
\proof

$\Rightarrow$

Assume $f$ is \strongcomp. Let $F \in \mathcal{RI}$  and $T$ be a Turing machine (an algorithm) computing the partition in Definition~\ref{def:fun-eps-comp-new}. We define the function-oracle Turing machine  $M$ as follows,
\begin{itemize}
  \item on input we have $n \in \nat$ and an oracle $\phi \in CF_x$

  \item we call $T$ with parameter $\eps=2^{-(n+2)}$ to obtain a sequence $\{I_i\}_{i=1}^L$, such that $K \subset \bigcup_{i=1}^L I_i$, $ f(I_i)\subset F(\eps, I_i)\wedge
\diam(F(\eps, I_i)) < \eps$

  \item let $\delta=\min_{i=1, \ldots, L} \diam(I_i)$ (by Lemma~\ref{lem:remove-zero-len} we can assume that we do not have intervals with zero length)
  \item let $m$ be the smallest such that $2^{-m} \leqslant \delta$,
  \item return any dyadic number $d$ from $F(\eps,I_j)$, where $I_j$ is any interval containing $\phi(m)$.
\end{itemize}

 To complete the proof we need to show that $|f(x) - d| < 2^{-n}$.
Since $|\phi(m)-x| \leqslant 2^{-m} \leqslant \delta $, then by Lemma~\ref{lem:cover} it follows that there exists $k_1,k_2$ such that $\phi(m) \in I_{k_1}$, $x \in I_{k_2}$
and $I_{k_1} \cap I_{k_2} \neq \emptyset$. Let us take any $y \in I_{k_1} \cap I_{k_2}$. We have
\begin{eqnarray*}
 |f(x)-d| \leqslant |f(x)-f(y)| + |f(y)-f(\phi(m))| + |f(\phi(m))-d| \leqslant \\
  \diam(F(\eps,I_{k_2})) +  \diam(F(\eps,I_{k_1}))+  \diam(F(\eps,I_{j})) \leqslant 3 \cdot 2^{-(n+2)} < 2^{-n}.
\end{eqnarray*}

$\Leftarrow$

Let us assume that $f$ is computable in the sense of Ko, thus by Theorem~2.13 in \cite{KiK91} $f$ has a recursive modulus function
$m$ on $K$. It means that
the function $m\colon \nat \to \nat$ such that for all $k\in\nat$ and all $x, y\in [a, b]$
\begin{equation}
|x - y| \leqslant 2^{-m(k)} \Longrightarrow |f(x) - f(y)| < 2^{-k}
\end{equation}
is computable. For any $\eps > 0$, if we partition $K$ into $I_1, \ldots, I_n$ intervals, where $n = \diam(K)\cdot 2^{m(\log_2 1/\eps)}$, then $\diam(I_i) = 2^{-m(\log_2 1/\eps)}$ and $\diam(f(I_i)) < 2^{-n} = \eps$,  for all $i = 1, \ldots, n$.
\qed

\subsection{Weak interval arithmetic computable (\weakcomp) functions}
\label{subsec:weak-comp}

We define a natural extension of the \strongcomp\ class. We allow countably (possibly infinitely) many representable intervals covering the domain of a function. Functions in \weakcomp\ are computable, but \weakcomp\ is a strict superset of \strongcomp. The ideas presented in  Grzegorczyk~\cite{Grzegorczyk55,Grzegorczyk57}, Ko~\cite{KF82}, Pour-El, Caldwell~\cite{E-PC75} and Shepherdson~\cite{Sh76} are equivalent to \strongcomp\ computability (cf.\ Theorem~\ref{thm:sc-equiv-ko}), hence they are valid in our framework and constitute an important subclass of \weakcomp\ functions.

\begin{definition}\label{def:fun-oldtype-eps-comp-new}
Let $f:\reals \supset \dom(f) \to\reals$ be such that its domain is a sum (possibly infinite) of representable intervals.

Let us fix $\eps>0$. The function $f$ is {\em $\eps$\weakcomp} if there exists $ F \in \mathcal{RI}$ such that there is an algorithm to establish a sequence (potentially infinite) of representable intervals $I_1,I_2,\ldots$ such that
\begin{equation}\label{eqn:e-weak-comp}
  \dom(f) = \bigcup_{i\in\nat} I_i  \quad\mbox{and}\quad \forall i :\ \left[ f(I_i)\subset F( I_i)\wedge
\diam(F( I_i)) \leqslant \eps\right].
\end{equation}

The function $f$  is {\em \weakcomp} if there exists $ F \in \mathcal{RI}$ such that for any given $\eps>0$  there is an algorithm to establish a sequence (potentially infinite) of representable intervals $I_1,I_2,\ldots$ such that

\begin{equation}\label{eqn:weak-comp}
\dom(f) = \bigcup_{i\in\nat} I_i   \quad\mbox{and}\quad \forall i:\ \left[ f(I_i)\subset F(\eps, I_i)\wedge
\diam(F(\eps, I_i)) \leqslant \eps\right].
\end{equation}

\end{definition}
Note that, if we omit pathological functions as for example
$$
f(x) = \left\{
\begin{array}{ll}
\omega, & x\in R\\
1, & x\in \reals\setminus R,
\end{array}
\right.
$$
where $\omega$ denotes an undefined value,
the process of choosing $I$ is algorithmic.

The following theorem says that a function $f$ is ($\eps$)\weakcomp, if there exists an algorithm $F$ (i.e.\ a program of an interval machine) that computes $f$ with requested precision. We require that for any (even non-computable) $x\in\dom(f)$ there exists an interval $I$, for which the algorithm $F$ returns a value suitably close to $f(x)$.

\begin{theorem}\label{thm:equiv-def}
Let us fix $\eps>0$. We say that a function $f:\reals \supset \dom(f) \to\reals$ is {\em $\eps$\weakcomp} iff there exists $ F \in \mathcal{RI}$ such that
\begin{equation}
  \forall x\in \dom(f)\ \exists I \in \irepreals, \ I \subset \dom(f) :\ x\in I,\ f(I)\subset F(I),\
\diam(F(I)) \leqslant \eps.\label{eqn:e-oldtype-comp-new}
\end{equation}
We say that $f$ is \weakcomp\   iff there exists $ F \in \mathcal{RI}$  such that
\begin{equation}
\forall \eps>0\  \forall x\in \dom(f)\ \exists I \in \irepreals, \ I \subset \dom(f)  :\ x\in I,\ f(I)\subset F(\eps,I),\
\diam(F(\eps, I)) \leqslant \eps.\label{eqn:comp-oldtype-new}
\end{equation}
\end{theorem}

\proof
The idea of the proof is exactly the same in both cases (computability and $\eps$-computability), thus we write $F([\eps,] I)$ to emphasize that the first parameter of $F$ is optional.

($\Rightarrow$)\\
We assume that $f$ is ($\eps$)\weakcomp, i.e.\ satisfies~(\ref{eqn:e-weak-comp}) or~(\ref{eqn:weak-comp}). Let us take any $x\in\dom(f)$. Since $\dom(f)$ is a sum of representable intervals $I_1, I_2, \ldots$, the number $x$ belongs to some $I_i$ and by (\ref{eqn:e-weak-comp}) (or~(\ref{eqn:weak-comp})) we know that the required $I\in\irepreals$ exists.

($\Leftarrow$)\\
We assume (in Remark~\ref{rem-countable}) that there are countably many representable numbers, hence there are countably many representable intervals (i.e.\ intervals with representable endpoints). To obtain countable covering of  $\dom(f)$ we inspect every representable interval (using for example primitive recursive Cantor pairing function) checking if it is contained in  $\dom(f)$  and $\diam(F([\eps,] I)) < \eps$. By the assumption that $f$
satisfies~(\ref{eqn:e-oldtype-comp-new}) or~(\ref{eqn:comp-oldtype-new}),
we know that for any $x\in \dom(f)$ the proper representable interval exists, thus we will find it in a finite number of steps.

Intervals that satisfy the requirements form the desired covering of $\dom(f)$.
\qed

Note that a remark analogous to Remark~\ref{rem:multi-dim-def1} can be made in this case.

\begin{theorem}
\label{thm:comp-cont-nonrep}
Assume $f:\reals\supset \dom(f)\to \reals$. If $f$ is \weakcomp\ then it is continuous at every non-representable point.
\end{theorem}
\proof
As $f$ is \weakcomp, by Theorem~\ref{thm:equiv-def}, there exists the corresponding $F\in \mathcal{RI}$.
Let us fix $\eps_0 > 0$ and $x_0\in\dom(f)$; now there exists $I_0\in\irepreals$ such that $x_0\in I_0$, $I_0 \subset \dom(f)$, $f(I_0)\subset F(\eps_0,I_0)$ and
$\diam(F(\eps_0, I_0)) < \eps_0$.

If $x_0\in\reals\setminus R$, then $x_0\in \mbox{int}(I_0)$ (recall that $\irepreals$ is the set of intervals with representable endpoints), hence $\exists \delta>0: B(x_0, \delta)\subset I_0$, where $B(x_0, \delta)$ is an open ball with a center at $x_0$ and radius $\delta$. Now, $\forall x:\ |x - x_0| < \delta \Rightarrow x\in I_0$, thus
\begin{eqnarray*}
|f(x) - f(x_0)| & \leqslant & \diam(f(I_0)) \\
  & \leqslant & \diam(F(\eps, I))\\
  & \leqslant & \eps.
\end{eqnarray*}
Hence, $f$ is continuous at $x_0$.
\qed

Below we consider a family of step functions $s_c$ and we prove that if $c$ is representable, then $s_c$, although discontinuous, is \weakcomp.

\begin{ex}\label{ex:f-obl}\rm
Consider a step function
$s_c(x) = [\ \mbox{if}\ (x\leqslant c)\ \mbox{then}\ 1\ \mbox{else}\ 2\ ]$.
If $c$ is a representable number then  $[c,c] \in \irepreals$.
Thus for $s_c$ there exists an interval extension $F \in \mathcal{RI}$, such that $\diam F(I) < \eps$:
$$F(I) = [\mbox{if } (I\leqslant [c,c]) \mbox{ then } [1,1]
\mbox{ else } \{
\mbox{if } (I > [c,c])\ \mbox{then}\ [2,2]
\mbox{ else }\ [1,2]\}].
$$
For $I \in \irepreals$ with $I\leqslant [c,c]$ or $I > [c,c]$, we have $\diam F(I)=0$.

Assume that $c$ is not a representable number.  Then any interval $I$ such that $c\in I$ has a form $[\alpha, \beta]$, where $\alpha < c < \beta$. Thus for any interval extension $F$ of the function $s_c$ and for any $I$ such that $c\in I$, $s_c(I) = \{1,2\}\subset F(I)$. Hence $\diam F(I)\geqslant 1 > \eps$.

Therefore, for $\eps < 1$, $s_c$ is $\eps$-\weakcomp\ if and only if  $c$ is representable.
\hfill\manual\char'170
\end{ex}

The following example is a \weakcomp\ function which has  infinitely many discontinuity points; it is  discontinuous for all $x \in \mathbb{D} \setminus \{0\}$.
\begin{ex}\label{ex:infinite}\rm
Let us consider a function $f: (0,1) \to \reals$:
$$
f(x)  = \left\{
\begin{array}{lp{5cm}}
\frac{1}{2^k}, & $x = \frac{m}{2^k}$ is an irreducible fraction \\
0, & otherwise.
\end{array}
\right.
$$

Assume that we know an algorithm (which is obvious to write) computing $\max_{x \in \mathbb{D} \cap I}(f(x))$. The function $f$ is \weakcomp\ since its interval extension is accomplished by $F\in \mathcal{RI}$:
\begin{align*}
F(I) = & [\mbox{if } (\mbox{left}(I)\neq \mbox{right}(I)) \mbox{ then } [0,\max_{x \in \mathbb{D} \cap I}(f(x))]\\
& \mbox{ else }
\{
\mbox{if } (\mbox{left}(I), \mbox{right}(I)\in\mathbb{D})\ \mbox{then}\ f(\mbox{right}(I))
\mbox{ else }\ [0,0]\}].
\end{align*}
\hfill\manual\char'170
\end{ex}

The above examples show that the class of \weakcomp\ functions  does not coincide with the ones
computable in the sense of Ko (Definition~\ref{def:comp-func-ko}) or Grzegorczyk (Definition~\ref{def:K-II}), which must be continuous.

The following remark explains/justifies that the set of \weakcomp\ functions in the sense of Definition~\ref{def:fun-oldtype-eps-comp-new} is not closed under composition.

\begin{remark}
Composition of two \weakcomp\ functions need not be a \weakcomp\ function. For example,
let $g(x) = \sin(x)$. Using the series expansion we can prove that $\sin(x)$ is \weakcomp.
Observe that function $s_0 \circ g$ is discontinuous at $\pi$, hence by Theorem~\ref{thm:comp-cont-nonrep} it cannot be \weakcomp.
\end{remark}

\subsection{Summary}

To summarize the properties of different definitions of interval arithmetic computability, let us remark:
\begin{enumerate}
\item $\eps$ interval arithmetic computability (i.e.\ $\eps$\weakcomp\ and $\eps$\strongcomp\ functions) results from the use of finite precision arithmetic; the non-$\eps$ variant is only possible when infinite precision arithmetic is used (with the only exception: a constant function);

\item interval arithmetic strong computability (i.e.\ $\eps$\strongcomp\ and \strongcomp\ functions)  implies the continuity of functions; non-strong variant allows discontinuous functions to be handled.

\end{enumerate}

\section{Existing models of computation}
\label{sec:othermodels}

There is no single definition of computability on real numbers.
What is computable depends on the assumed model of computation which in turn depends on our requirements or visions.
There are approaches that consider functions on all real numbers: real-RAM and bit models; on the other hand, the are settings (belonging to a field of constructive mathematics) which work on (specifically defined) computable numbers only: e.g. Banach/Mazur (cf.~\cite{SM63}) or Pour-El/Richards (cf.~\cite{PER}). We are interested in computations that allow the use of all real numbers and not just a computable subset thus we skip the latter constructive approach.
 We follow \cite{Bra05} and \cite{Wei00} in a short presentation of different models.

\subsection{Real-number models}
In real-number models the reals  are considered {\em real}, i.e.\ actual objects.
 The model of computation is a generalized Random Access Machine with two kinds of registers: for natural and for real numbers.

The two approaches we briefly touch upon, BSS and IBC, are examples of the real-number models.

\subsubsection{Blum-Shub-Smale (BSS) model}
\label{subsec:BSS}

In \cite{BCSS} the authors point to the incompatibility between the discrete world of computer computations and the continuous nature of calculus and numerical analysis. Since the discrete  computer  cannot cope with the continuous mathematics, they decide to change the computation model.
This idealization makes perfect sense in many contexts, as it allows to put the classical numerical analysis on a firm basis.
In the BSS approach arithmetic operations are performed on real numbers that are stored with infinite precision. It is obvious that no physical model implements this approach, however this is nearly true when manipulating polynomials or in a lot of tasks of numerical linear algebra.

In the BSS model the authors define  computable functions to be the ones produced by programs using the arithmetic operations and the inequalities for branching.  Therefore a domain of  computable function $f$ in the BSS model is  a possibly countable collection of semi-algebraic sets $\{A_i\}$ and $f$ is rational on $A_i$.

There are functions obviously non-computable on present computers, but computable in the BSS model. For example the function:
$$
f(x) = \left\{
\begin{array}{ll}
1, & x\in\rat\\
0, & x\not\in\rat \mbox{ and } x^2\in\rat\\
\omega, & \mbox{otherwise},
\end{array}
\right.
$$
where $\omega$ means``non-defined''. This means that program does not stop for such input.

On the other hand, there are functions computable with some error $\eps$, classified as non-computable in the BSS model; for example the square root or the exponential function. This is not a fundamental issue; it is just a matter of definition; see \cite{Bra05} where a suitable extension of definition of computability of the real function inside the BSS model is given.

Observe that, even in the context of the BSS model, for example when dealing with solutions of ODEs, the intervals might be needed to carry out the estimates. In that case according to the spirit of the BSS model the set of representable numbers to be used in the construction of the interval arithmetic should be $\mathbb{R}$ (see Section~\ref{sec:intervals} for the brief discussion of the interval arithmetic).

\subsubsection{The Information Based Complexity}
\label{subsubsec:ibc}
The focus of Information Based Complexity (IBC)~\cite{IBC1988, TW91, TW92} is on the complexity of the problems in continuous mathematics. There is no explicit need for the definition of  computable functions in the IBC. The lower bound for the complexity of some problems, for example the computation of a definite integral, is obtained by considering the following (optimization) problem: given the values of $f$ or its derivatives at $n$ points (evaluation nodes) find the radius of the ball containing all possible values of the integral and minimize this radius over all possible locations of the evaluation nodes. This gives a minimal error  $\eps(n)$, which any program must make if it  asks  just for $n$ values of function
or its derivatives at some points. The $n(\eps)$, which is inverse to $\eps(n)$, give then the lower bound for the complexity of computing the definite integral with the precision $\eps$. If then one can find an algorithm $\mathcal{A}$ using some oracles returning the values of the function or its derivatives, for which the cost matches the lower bound $n(\eps)$, then the claim can be made that the obtained function $n(\eps)$ is the complexity of the problem and  $\mathcal{A}$ is an optimal algorithm (under the given type of admissible  information). The cost of computation in the IBC depends on the computation model  {\em (...) the complexity of a problem can be totally different in real number and bit models}; cf.~\cite{TW91}.

The goals of IBC include the idenfication and design of (almost) optimal algorithms. This is exactly our goal, however
from the perspective of the interval machine the IBC approach makes typically unjustified assumption about the kind of information a computer program may ask for. As it was stressed in Section~\ref{sec:int-alg}, for the computable functions on the interval machine we can ask not only for the values of the functions or derivatives at some points, but also for the estimates of these values over some interval sets. But for this to make sense, one needs to define a notion of the computable function and computation model, which the IBC appears to deliberately avoid while striving for the largest possible generality and to be largely independent of the particular computation model assumed.

\subsection{Bit-models}

\subsubsection{Type-2 Theory of Effectivity (TTE)}

Type-2 Theory of Effectivity (TTE) described by K. Weihrauch in~\cite{Wei00} provides a general framework for investigating problems in computable analysis.

The basic TTE computability concepts are implemented via naming systems. There exists also an implementation in C++ of exact real arithmetics based on TTE - the iRRAM simulator by N. M\"uller  \cite{iRRAM}.
Although the TTE is more realistic then the BSS model, even this attempt diverges from (clashes/conflicts with CAPA) practice:
\begin{enumerate}
\item Our main objection is the infeasibility of TTE representations of mathematical objects (numbers,  functions etc) (both input and output) as infinite strings. Programs in real life do not generate output bit by bit; in order to get a new decimal digit on output, we rather need to run the program again with parameters forcing the better precision.

    This is manifestly a theoretical construct, because it computation in practice is finite. Therefore it appears to us  artificial.

\item Real computers operate on finite representations thus simple discontinuous functions as
$$
s_1(x) = \left\{
\begin{array}{ll}
1, & x\geqslant 1\\
0, & \mbox{\rm otherwise},
\end{array}
\right.
$$
are obviously computable from the CAPA  view point (see also our definition of computable functions in Section~\ref{sec:comp-func-intv-machine}), while the TTE classifies them as uncomputable.
\end{enumerate}

The notion of complexity in TTE is subtle. Good approaches to this isssue are Ko's model (discussed below) or the recent ones proposed in \cite{CookKawa, NeumannSteinberg}.

\subsubsection{Ko's approach }\label{sec:ko}

K.I.~Ko (\cite{KiK91, KF82}) applies ideas from the recursion theory and the classical complexity theory to real functions.  For certain representations of real numbers his approach is equivalent to Weihrauch's (see for example~\cite{Wei00}, Theorem~9.4.3). In fact, TTE is more general, however the  Ko's approach constitute a commonly accepted standard. There is a lot of results based on that model concerning problems in numerical analysis, in both aspects: computability and complexity. A recent example is the paper \cite{PoulyGraca-2016} on the complexity of solving ODEs.

%--------------------------------------

When discussing the computable real functions, Ko first introduces the idea of computable numbers, then the set of functions binary converging to $x$ and finally the computable functions; we quote the below:

\begin{definition}\label{def:ko-num}
We say a real number $x$ is {\em computable} if there is a computable function $\phi:\ \nat\to\mathbb{D}$ such that for all $n\in N$, $|\phi(n) - x|\leqslant 2^{-n}$.
\end{definition}

\begin{definition}\label{def:ko-fun-bin}
For each real number $x$, a function $\phi:\ \nat\to\mathbb{D}$ is said to {\em binary converge} to $x$ if it satisfies the condition that for
all $n\in\nat$, $\phi(n) \in \{m/2^{n},\ m \in \mathbb{Z}\}$ and $|\phi(n) - x|\leqslant 2^{-n} $. Let $\mbox{CF}_x$ (Cauchy function) denote the set of all functions binary converging to $x$.
\end{definition}

Intuitively, Turing machine $M$ with one oracle computes a real function $f$ in the following way:
\begin{enumerate}[(a)]
\item The input $x$ to $f$, represented by some $\phi\in \mbox{CF}_x$, is given to $M$ as an oracle.
\item The output precision $2^{-n}$ is given in the form of integer $n$
(or, in unary notation, a string $0^n$) as the input to $M$.
\item The computation of $M$ usually takes two steps, though sometimes these two steps may be repeated an indefinite number of
times:
\begin{enumerate}[(i)]
\item $M$ computes, from the output precision $2^{-n}$, the required input
precision $2^{-m}$,
\item $M$ queries the oracle to get $\phi(m)$, such that $|\phi(m) - x| \leqslant 2^{-m}$,
and computes from $\phi(m)$ an output $d\in\mathbb{D}$ with $|d - f(x)|\leqslant 2^{-n}$
\end{enumerate}
\end{enumerate}

The precise definition is as follows.
\begin{definition} \cite[Def. 2.11]{KiK91}
\label{def:comp-func-ko}
  A real function $f:\mathbb{R}\to \mathbb{R}$ is \emph{computable} if there is a function-oracle TM (Turing machine) $M$ such that for each $x \in \mathbb{R}$ and each $\phi \in CF_x$, the function $\psi$ computed by $M$ with oracle $\phi$ (i.e. $\psi(n)=M^\phi(n)$ is in $CF_{f(x)}$. We say that $f$ is \emph{computable on interval $[a,b]$} if the above condition holds for all $a \in [a,b]$.
\end{definition}

Note that the quantification ``for each $x \in \mathbb{R}$ and each $\phi \in CF_x$'' in Definition~\ref{def:comp-func-ko} is essential in the proof of the continuity and computability of the modulus function  of computable functions (see Theorem 2.13 in~\cite{KiK91}).

This proofs proceeds as follows:
Let us take an arbitrary $x\in\reals$. As the Cauchy function $\phi\in CF_x$ take $b_x$, the binary representation of $x$ truncated to $n$ binary places, i.e.\ $|\phi(n) - x| = |b_x(n) - x|\leqslant 2^{-n}$. Now, following the proof of Theorem 2.13, observe that for all $y$ such that $|y-b_x(k_x)|\leqslant 2^{-k_x}$
\begin{equation*}
  |f(y) - M^{b_x}(n+2)| \leqslant 2^{-(n+1)},
\end{equation*}
where
$k_x=\max\{k \ | \ b_x(k) \ \mbox{is queried in the computation $M^{b_x}(n+2)$}\}.$
This gives rise to an actual use of an oracle, which might indeed be uncomputable.

\begin{ex}\rm
Consider $x > 0$ such that the $n$-th digit of $x$ is
1, if the $n$-th Turing machine stops on $n$, and 0 otherwise. The binary expansion of $x$ is
$$
\psi_x(n) = \left\{
\begin{array}{ll}
1, & \mbox{if the $n$-th Turing machine stops on $n$}\\
0, & \mbox{otherwise}
\end{array}
\right.
$$
and the binary Cauchy representation is
$b_x(n) = \sum_{k = 1}^{n}\psi(k)\cdot 2^{-k}.$
It is clear that there is no Turing machine computing $b_x$.
\hfill\manual\char'170
\end{ex}

The oracle-Turing machine for the computable function (in the sense of Definition~\ref{def:comp-func-ko}) for given $n$ and binary representation of the number $x$ up to $m(n)$ binary places, returns $f(x)$ up to an error $2^{-n}$. Thus, in terms of intervals,  the machine performs two operations: seeks for the right value of $m(n)$ and then computes
\begin{equation*}\label{eq:Ko-comp-func-intervals}
f({\rm trunc}_n(x)+2^{-m(n)}[-1,1]) \subset f(x) + 2^{-n} \cdot [-1,1].
\end{equation*}
The question is how this inclusion is achieved?
In Section~\ref{sec:subdiv-eval} we discuss an algorithm for the interval machine (equipped with an interval arithmetic), which has a built-in mechanism to return a rigorous bound for $f(I)$, where $I$ is an interval (or product of intervals) and $f$ is an elementary function.

 The present paper defines a model of computation which should be used for the estimation of the complexity of the problems and the quality of the interval algorithms. The work on complexity in Ko's framework is quite extensive, but it is not obvious whether the results are relevant to CAPAs. A membership to a class of complexity (as P or PSPACE) is an important fact, but it does not give any insight into how an optimal algorithm could look, which we are most interested in. 

%--------------------------------------

\subsubsection{Definitions of Grzegorczyk}\label{sec:grzegorczyk}
In his paper~\cite{Grzegorczyk57} Grzegorczyk gives several equivalent definitions of real computable functions, which lead to the computability
concept equivalent to that of Ko. These are interesting from our view point as they avoid the use of un-computable oracles. We present two of them below.

Grzegorczyk starts from definitions of computable real number and sequence, based on the definition of the class $\mathcal{K}$ of computable functionals; see~\cite{Grzegorczyk55}.

\begin{definition}
A {\em real number $a$ is computable ($a\in \mathcal{K}$)} if there exists a recursive function $f:\nat\to\nat$ such that
\begin{equation}\label{def:grzeg-num}
\left|a - \frac{f(n)}{n+1} \right| < \frac{1}{n+1}, \quad\forall n\in\nat.
\end{equation}

A {\em real sequence $\{a_n\}$ is computable ($\{a_n\}\in \mathcal{K}$)} if there exists a recursive function $g:\nat^2\to \nat$ such that for all $k\in \nat$
\begin{equation}\label{def:grzeg-seq}
\left|a_n - \frac{g(n, k)}{k+1} \right| < \frac{1}{k+1}, \quad \forall n\in \nat.
\end{equation}

A {\em real function $\varphi$ is computable ($\varphi\in \mathcal{K}_I$)} if and only if there exists functional $\Phi\in\mathcal{K}$ such that for all $a\in \reals$ and for all $h:\nat\to\nat$
\begin{equation}\label{def:grzeg-fun}
\mbox{if } \left|\frac{h(k)}{k+1} - a \right| < \frac{1}{k+1}, \quad \forall k\in\nat,\quad \mbox{ then } \left| \frac{\Phi(h)(k)}{k+1} - \varphi(a)\right| < \frac{1}{k+1},\quad \forall k\in\nat.
\end{equation}
\end{definition}

Note that, in the above definitions functions $f$ and $g$ are recursive, but $h$ not need to be computable, therefore it plays the role of an oracle.

\begin{definition}\label{def:K-II}
$\varphi\in\mathcal{K}_{II}$ if and only if the following two conditions are satisfied:
\begin{enumerate}
\item For each computable sequence $\{a_n\}$, the sequence of values $\{\varphi(a_n)\}$ is also computable.
\item $\varphi$ is computable uniformly continuous with respect to the rational segments. This means that there exists a recursive function $g:\nat^3\to\nat$, such that
\begin{equation*}
\mbox{if } r_n < a,b < r_m \mbox{ and } |a - b| < \frac{1}{g(n,m,k)}\quad \mbox{ then } \left| \varphi(a) - \varphi(b)\right| < \frac{1}{k+1},
\end{equation*}
$\forall n,m,k\in\nat$ and $a,b\in\reals$, where $r_n, r_m\in\rat$ are $n$-th and $m$-th rational number, respectively.
\end{enumerate}
\end{definition}

The Definition~\ref{def:K-II} of computable real functions, though equivalent to Ko's, does not use any oracles. Nonetheless, we think that for interval computations the definition expressed in terms of intervals is more natural.

\section{Conclusions}
In the paper we present the idea of interval computability which has some advantages over other models of real computation:
\begin{enumerate}
\item it operates on finitely represented numbers being the ends of intervals;
\item it allows to obtain interval information, i.e.\ values of functions not only at a point, but also on the whole interval;
\item it allows rigorous interval calculations (which are crucial for CAPAs) giving the certainty that all real numbers (whether computable or not) in the interval possess a property in question;
\item it has an implementation in the form of various packages for rigorous numerics (e.g.\ CAPD, see~\cite{CAPD});
\item it is a straightforward tool suitable for laymen in the field of computational analysis.
\end{enumerate}
\section*{References}
\bibliographystyle{plain}
\bibliography{ref}

\begin{appendices}

\section{Basic terms and concepts of interval arithmetic}
\label{sec:intervals}

Let us recall that by ${\irepreals}$ we denote the set of all intervals with endpoints belonging to $\widehat{R}$.
Some of the notations and definitions related to intervals are taken from the book~\cite{MKC09}. In particular, we use capital letters to denote intervals and their endpoints are marked with a line below or above the letter: $X = [\underline{X},\overline{X}]$.
Two intervals $X$ and $Y$ are said to be equal, if they are the same sets i.e.
$$X = Y \Longleftrightarrow
\left(\underline{X} = \underline{Y} \wedge \overline{X} = \overline{Y}\right).$$

We say that $X$ is {\em degenerate} if $\underline{X} = \overline{X}$. Such an interval contains a single representable number $x$.
By convention, we identify a degenerate interval $[x, x]$ with the representable number $x$. In
this sense, we may write such equations as
$0 = [0, 0]$.

\begin{definition}\label{def:pred}
The class of interval predicates $\mathcal{C}$ consists of:
\begin{itemize}
\item interval equality and inequalities:
\begin{eqnarray*}
X =Y & \Longleftrightarrow & \underline{X} = \underline{Y}\wedge \overline{X} = \overline{Y}\\
X<Y & \Longleftrightarrow & \overline{X} < \underline{Y}
\end{eqnarray*}
\item interval set inclusion:
\begin{equation}
X \subset Y \Longleftrightarrow (\underline{Y} \leqslant \underline{X} \wedge \overline{X} \leqslant \overline{Y}) .
\end{equation}
\end{itemize}
\end{definition}

Observe that when applied to interval arguments it might happen that neither $x \leqslant y$ nor  $y > x$ holds.
We also have that if $X < Y$, then for all $x\in X$ and for all $y \in Y$, $x < y$. However from  $X=Y$ we cannot infer that $x=y$ for all $x \in X$ and $y \in Y$.
This has  consequences for programming using interval arithmetic.

\begin{definition}\label{def:arith-ext}
We define {\em interval extensions} of arithmetic operations (for division we assume that $0\not\in Y$):
\begin{eqnarray*}
X + Y & = & [\underline{X}+\underline{Y}, \overline{X}+\overline{Y}]\\
X-Y & = & [\underline{X}-\underline{Y}, \overline{X}-\overline{Y}]\\
X\cdot Y & = & [\min\{\underline{X}\cdot\underline{Y}, \underline{X}\cdot\overline{Y}, \overline{X}\cdot\underline{Y}, \overline{X}\cdot\overline{Y}\}, \max\{\underline{X}\cdot\underline{Y}, \underline{X}\cdot\overline{Y}, \overline{X}\cdot\underline{Y}, \overline{X}\cdot\overline{Y}\}]\\
X/Y & = & X\cdot [1/\overline{Y}, 1/\underline{Y}]\\
 & = & [\min\{\underline{X}/\underline{Y}, \underline{X}/\overline{Y}, \overline{X}/\underline{Y}, \overline{X}/\overline{Y}\}, \max\{\underline{X}/\underline{Y}, \underline{X}/\overline{Y}, \overline{X}/\underline{Y}, \overline{X}/\overline{Y}\}],
\end{eqnarray*}
\end{definition}

Note that interval operations defined above do not take into account that, even if the end points are representable numbers, the results of the arithmetic operation might not be  representable (this is the case of the floating point arithmetic present on today's machines). As mentioned in Section~\ref{subsec:nat-ph}, we require that actual results belong to the computed intervals;
to achieve this we require that  computer implementation $\widehat{\diamond}$ of a theoretical interval operation $\diamond$ satisfies
\begin{equation}\label{eqn:int-op}
 \forall X,Y \in \irepreals \quad   X \diamond Y \subseteq X\ \widehat{\diamond}\ Y \in \irepreals, \quad \mbox{where} \quad \diamond \in \{+,-, \cdot, / \}.
\end{equation}
and for division we assume that $0\not\in Y$.

\begin{definition}\label{def:unary-op}
We define the {\em unary basic operations}:
\begin{enumerate}
\item
the {\em left} endpoint of an interval $X$ is given by
$\mbox{\rm left}(X) = [\underline{X}, \underline{X}] \equiv\underline{X}.
$

\item the {\em right} endpoint of $X\in {\irepreals}$ is given by
$\mbox{\rm right}(X) = \overline{X}.$

\item the {\em absolute value} of $X\in {\irepreals}$, denoted $|X|$, is the maximum of the absolute values of its
endpoints:
$|X| = \max\{|\underline{X}|, |\overline{X}|\}.$
Note that $|x| \leqslant |X|$ for every $x \in X$.

\end{enumerate}
\end{definition}
 Note that for $X\in{\irepreals}$ basic unary operations are functions ${\irepreals}\to {\irepreals}$.

\subsection{Reasonable arithmetic}

 If we do not state any requirement for the quality of interval arithmetic, it can give arbitrarily large error in the implementation of real functions.
Thus we discard unreasonable implementations and we want to be as realistic as possible.
We require the computer implementation to be reasonable, i.e.\ $\diam(X\ \widehat{\diamond}\ Y)$ to be as small as can possibly be obtained.
In floating point arithmetic this is achieved by an application of  the directed rounding when performing the non-rigorous arithmetic operations, i.e.\ when computing the left end of an interval the numbers are rounded down, whilst for the right end they are rounded up.

\begin{definition}
\label{def:reasonable-int-arth}
The computer implementation is {\em reasonable}, if for finite $R$, the result of an operation $X\diamond Y$ is an interval with left endpoint being the greatest representable number less than or equal to $\underline{X\diamond Y}$ and right endpoint the least representable number greater than or equal to $\overline{X\diamond Y}$.
If $R = \rat$, then $X\diamond Y = X\ \widehat{\diamond}\ Y$.
\end{definition}

\begin{definition}\label{def:int-art}
 A set of  reasonable computer implementations of interval extensions of $\{+, -, \cdot, /\}$ is called an {\em interval arithmetic}.  If $R = \rat$, we say that we have the interval arithmetic with infinite precision.
\end{definition}

\section{Model of an interval machine}\label{def:machine}

When a CAPA  is performed, we deal with the really existing computers and no idealizations are made, beside believing
that the processor and compilers work as promised in the documentation. For this reason it might be
worth to formalize the model of a machine, which will correspond to the reality of CAPAs.
We call it an \emph{interval machine}.  It is equipped with arithmetic which operates on intervals as described in Section~\ref{sec:intervals}.

{\em An interval machine} $M$ over
representations $\widehat{R}\subset \Sigma^*$, where $\Sigma$ is a fixed, finite alphabet, is a program that can be
presented as a flow diagram (a finite connected graph) with nodes representing instructions. Instructions are described in Section~\ref{subsec:nodes} and the execution of the program in Section~\ref{subsec:comp}.

\subsection{A program of an interval machine}\label{subsec:nodes}

A program contains nodes of the following form:
\begin{enumerate}
\item  Exactly one {\em input node} {\sc start} with input variables $X_1, \ldots, X_n$ of type $\irepreals$ (i.e.\  variables store representable intervals) and exactly one {\em output node} {\sc stop}, where we specify  intervals returned as shown in Figure~\ref{node:in-out}.
An input node has one outgoing edge, and no incoming edges; an output node has no outgoing edges, but at least one incoming edge.
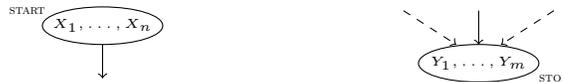
\begin{figure}[htb]
\centering
\begin{tikzpicture}
\node[] at (0,1.2) {\tiny\sc start};
\draw (1.0,1.0) ellipse (0.8cm and 0.25cm) node[] {\tiny $X_1, \ldots, X_n$};

\draw[->] (1.0,0.75) -- (1.0,0.3);

\draw (6.0,0.5) ellipse (0.8cm and 0.25cm) node[] {\tiny $Y_1, \ldots, Y_m$};

\draw[->] (6.0,1.2) -- (6.0,0.75);

\draw[dashed,->] (5.0,1.2) -- (5.7,0.75);

\draw[dashed,->] (7.0,1.2) -- (6.3,0.75);

\node[] at (7.0,0.3) {\tiny\sc stop};
\end{tikzpicture}
\caption{An input node and an output node with output values $Y_1, \ldots, Y_m \in\irepreals$ }\label{node:in-out}
\end{figure}

\item {\em Assignment nodes}  as shown in Figure~\ref{node:asg}.

The right hand side can be a constant $[c_1, c_2]$ or one of the following basic operations (denoted by $F(X)$ in Figure~\ref{node:asg}):
\begin{itemize}
\item any basic interval arithmetic operations (see Definition~\ref{def:arith-ext}),

\item any of the unary basic operation (see Definition~\ref{def:unary-op}),
\item a variable,
\item a value popped from a stack (see description of stack operations at~\ref{def:stack-op}).
\end{itemize}

In an instruction $Z\leftarrow [c_1, c_2]$ we assume that both $c_1, c_2$ are representable numbers, where $c_1\leqslant c_2$.
On the left hand side, $Z$ can be a new variable or a previously existing one.
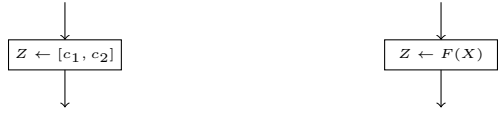
\begin{figure}[ht]
\centering
\begin{tikzpicture}

\draw[->] (1.0,1.0) -- (1.0,0.5);

\draw (0.25,0.1) rectangle (1.75,0.5) node[midway] {\tiny $Z\leftarrow [c_1, c_2]$};

\draw[->] (1.0,0.1) -- (1.0,-0.4);

\draw[->] (6.0,1.0) -- (6.0,0.5);

\draw (5.25,0.1) rectangle (6.75,0.5) node[midway] {\tiny $Z\leftarrow F(X)$};

\draw[->] (6.0,0.1) -- (6.0,-0.4);

\end{tikzpicture}
\caption{Assignment nodes.}\label{node:asg}
\end{figure}

\item {\em Branching nodes}  as shown in Figure~\ref{node:test}.
The predicate $C$ is one of the interval predicates (see Definition~\ref{def:pred}).
If $C(Y)$ is satisfied, then we use a branch {\sc yes}; if it is not satisfied, we use a branch {\sc no}; if it is not defined, the result of the operation is not defined. Note that $Y$ can be a vector of variables.
\begin{figure}[htb]
\centering
\begin{tikzpicture}
\draw[->] (1.95, 1.7) -- (1.95,1.2);

\draw[->] (0.6,1) -- (0.6,0.5);
\draw (0.6,1) -- (1,1) node[anchor=south east] {\tiny\sc yes};

\draw (1,1) -- (1.2, 1.2);
\draw (1,1) -- (1.2, 0.8);
\draw (1.2, 1.2) -- (2.2, 1.2);
\draw (1.2, 0.8) -- (2.2, 0.8) node[midway, above] {\tiny $C(Y)$};
\draw (2.2, 1.2) -- (2.4,1.0);
\draw (2.2, 0.8) -- (2.4,1.0);

\draw (2.4,1.0) -- (2.8,1) node[anchor=south east] {\tiny\sc no};
\draw[->] (2.8, 1.0) -- (2.8, 0.5);

\end{tikzpicture}
\caption{A branching node}\label{node:test}
\end{figure}
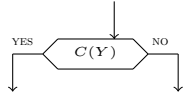

\item\label{def:stack-op} {\em Stack operations} as shown in Figure~\ref{node:stack}. There are three kinds of stack operations:
\begin{itemize}
\item {\sc empty}$(S)$ -- checking if $S$ is an empty stack; if $S$ does not exist, then the result of the operation is undefined,
\item {\sc push}$(S, X)$ -- push a value $X$ onto a stack $S$; if stack $S$ does not exist, a new one is created,
\item {\sc pop}$(S)$ -- returns a value form the top of a stack $S$; it removes the returned value from the top. This has to be a right hand side of an assignment operation. If stack $S$ is empty or it does not exist, then the result of the operation is undefined.
\end{itemize}
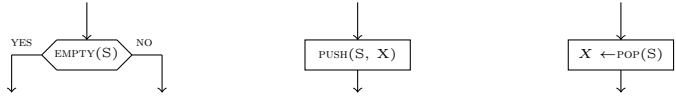
\begin{figure}[htb]
\centering
\begin{tikzpicture}
\draw[->] (0.6,1) -- (0.6,0.5);
\draw (0.6,1) -- (1,1) node[anchor=south east] {\tiny\sc yes};

\draw (1,1) -- (1.2, 1.2);
\draw (1,1) -- (1.2, 0.8);
\draw (1.2, 1.2) -- (2.0, 1.2);
\draw (1.2, 0.8) -- (2.0, 0.8) node[midway, above] {\tiny {\sc empty}(S)};
\draw (2.0, 1.2) -- (2.2,1.0);
\draw (2.0, 0.8) -- (2.2,1.0);

\draw (2.2,1.0) -- (2.6,1) node[anchor=south east] {\tiny\sc no};
\draw[->] (2.6, 1.0) -- (2.6, 0.5);

\draw[->] (1.6, 1.7) -- (1.6,1.2);

\draw (4.5,0.8) rectangle (5.9,1.2) node[midway] {\tiny {\sc push}(S, X)};
\draw[->] (5.2, 1.7) -- (5.2,1.2);
\draw[->] (5.2, 0.8) -- (5.2,0.5);

\draw (8.0,0.8) rectangle (9.4,1.2) node[midway] {\tiny $X\leftarrow ${\sc pop}(S)};
\draw[->] (8.7, 1.7) -- (8.7,1.2);
\draw[->] (8.7, 0.8) -- (8.7,0.5);

\end{tikzpicture}
\caption{Stack operations}\label{node:stack}
\end{figure}

\end{enumerate}

A notation $Z\leftarrow F(X)$ is also an abbreviation of the whole block of instructions computing the function $F$, assuming that $F$ can indeed be computed with elementary instructions of the machine.

For the sake of clarity, we adopt a convention that a stack and a variable cannot share a name.

In any situation not described above, the operation of the machine is undefined.

Internal memory is allocated for variables, while the external memory is allocated for the implementation of stacks.
There can be any finite number of variables and stacks used in the machine.

\subsection{Computations on an interval machine}
\label{subsec:comp}

We assume that on input data $X_1, \ldots, X_n \in\irepreals$ machine $M$ executes a program with sequence of instructions (nodes) $\{P_0, P_1, \ldots, P_m, P_{m+1}\}$.
\begin{enumerate}
\item {\bf Initial state: } The program starts at the $P_0$ = {\sc start} with
input data $X_1, \ldots, X_n$.

\item {\bf Intermediate state: } The program executes consequently instructions from $\{P_1, \ldots, P_m\}$. If current instruction is $P_i$, then it modifies variables and stacks according to that instruction and proceeds to the next instruction $P_{i+1}$.

\item {\bf Final state: } The program ends with $P_{m+1}$ = {\sc stop} and the output $Y_1, \ldots, Y_k \in\irepreals$.

\end{enumerate}

\begin{definition}\label{def:int-machine-comp}
We say that {\em an interval machine $M$ computes a function $F:\ {\cal X}^n\to {\cal X}^k$} if for any $(X_1, \ldots, X_n)\in \dom(f)$ starting at the initial state with an input data $X_1, \ldots, X_n$, it arrives at the final state
and returns $(Y_1, \ldots, Y_k) \in \irepreals^k$, such that $(Y_1, \ldots Y_k) = F(X_1, \ldots, X_n)$.
\end{definition}

\begin{ex}\rm
Let us consider a function
$$f(x,y) = [\ \mbox{if}\ (x+y\leqslant 0)\ \mbox{then}\ 0\ \mbox{else}\ 1\ ]$$
An interval extension of $f$ is a function
$$F(I_x, I_y) \equiv [\mbox{if } (I_x+I_y\leqslant [0,0]) \mbox{ then } [0,0]
\mbox{ else } \{
\mbox{if } (I > [0,0])\ \mbox{then}\ [1,1]
\mbox{ else }\ [0,1]\}].
$$
A program of an interval machine computing $F$ is presented on Figure~\ref{fig:int-mach}

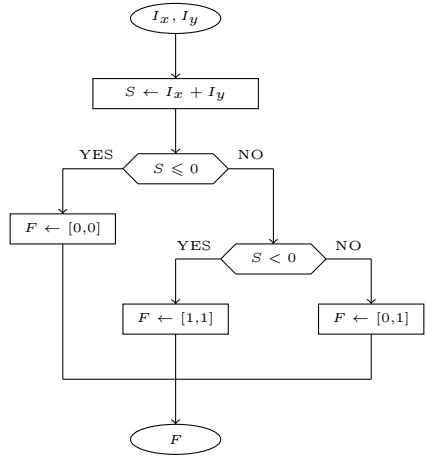
\begin{figure}[htb]
\centering
\begin{tikzpicture}
\draw (5.1,5.6) ellipse (0.6cm and 0.2cm) node[] {\tiny  $F$};

\draw (3.6, 6.4) -- (7.7, 6.4);
\draw[->] (5.1, 6.4) -- (5.1, 5.8);

\draw (3.6, 8.2) -- (3.6, 6.4);
\draw (7.7, 7.0) -- (7.7, 6.4);
\draw (5.1, 7.0) -- (5.1, 6.4);

\draw[->] (6.4,9.2) -- (6.4, 8.2);
\draw (5.7, 8.0) -- (5.9, 8.2);
\draw (5.7, 8.0) -- (5.9, 7.8);
\draw (5.9, 8.2) -- (6.9,8.2) node[midway, below] {\tiny $S < 0$};
\draw (5.9, 7.8) -- (6.9,7.8);
\draw (6.9, 8.2) -- (7.1, 8.0);
\draw (6.9, 7.8) -- (7.1, 8.0);
\draw (7.1, 8.0) -- (7.7, 8.0) node[anchor=south east] {\tiny NO};
\draw[->] (7.7,8.0) -- (7.7, 7.4);
\draw (7.0, 7.0) rectangle (8.4, 7.4) node[midway] {\tiny  $F \leftarrow$ [0,1]};

\draw (5.1, 8.0) -- (5.7, 8.0) node[anchor=south east] {\tiny YES};
\draw[->] (5.1,8.0) -- (5.1, 7.4);
\draw (4.4, 7.0) rectangle (5.8, 7.4) node[midway] {\tiny  $F \leftarrow$ [1,1]};

\draw (4.4, 9.2) -- (4.6, 9.4);
\draw (4.4, 9.2) -- (4.6, 9.0);
\draw (4.6, 9.4) -- (5.6,9.4) node[midway, below] {\tiny $S \leqslant 0$};
\draw (4.6, 9.0) -- (5.6,9.0);
\draw (5.6, 9.4) -- (5.8, 9.2);
\draw (5.6, 9.0) -- (5.8, 9.2);
\draw (5.8, 9.2) -- (6.4, 9.2) node[anchor=south east] {\tiny NO};
\draw (3.6, 9.2) -- (4.4, 9.2) node[anchor=south east] {\tiny YES};
\draw[->] (3.6,9.2) -- (3.6, 8.6);
\draw (2.9, 8.2) rectangle (4.3, 8.6) node[midway] {\tiny  $F \leftarrow$ [0,0]};

\draw[->] (5.1, 10.0) -- (5.1, 9.4);
\draw (4.0,10.0) rectangle (6.2,10.4) node[midway] {\tiny $S\leftarrow I_x + I_y$ };

\draw[->] (5.1, 11.0) -- (5.1, 10.4);
\draw (5.1,11.2) ellipse (0.6cm and 0.2cm) node[] {\tiny  $I_x, I_y$};
\end{tikzpicture}

\caption{Diagram computing $F(I_x, I_y)$}\label{fig:int-mach}

\end{figure}

\end{ex}

\subsection{Statements about the power of the interval machine}
Notice that the model of interval machine is ideologically very similar to a Random Access Machine in its traditional form, which is Turing equivalent. Hence the next two remarks (we omit the proofs as the results are fairly intuitive). The conclusion from these remarks is that an interval machine does not introduce any new operations. However, it has the advantage of specific notation tailored to the interval arithmetic.

\begin{remark}
If $F$ is a function computed by an interval machine $M_F$, then there exists a deterministic Turing machine $T_F$ computing $F$.
\end{remark}

\begin{remark} \label{rem:equiv2}
For any RAM machine computing a function, there exists an interval machine computing the same function, where the integer
  values $n$ are represented as degenerate intervals $[n,n]$.
\end{remark}

The power of the interval machine from the point of view of CAPA is summarized by the following statements:
\begin{enumerate}
\item \emph{Let $f:\reals^n\supset \dom(f)\to \reals^m$
be a rational function (i.e.\ the function that can be represented as a straight-line program evaluating $f$ using interval arithmetic) and let $K \subset \dom(f)$ be a compact set that can be covered by cubes of arbitrarily small diameter with representable endpoints.
Then  $\sup_{x \in K} f(x)$ can be estimated
from above with arbitrary accuracy} (for the one-dimensional case see Remark~\ref{rem:BSA-inf}).

\item
\emph{Every mathematical theorem which can be reduced to a finite number of strict
inequalities between rational functions on some simple (``computable'') compact in $\mathbb{R}^n$
can be proved as CAPA on the interval machine.}
\end{enumerate}

The converse of the second statement is obviously not true; see for example the symbolic computations leading to identities about trigonometric functions. 
\end{appendices}

\end{document}